# Multiscale Mechanical Consequences of Ocean Acidification for Cold-Water Corals

Uwe Wolfram[1*]  Marta Peña Fernández[1]  Samuel McPhee[1]  Ewan Smith[1]  Rainer J. Beck[1]  Jonathan D. Shephard[1]  Ali Ozel[1]  Craig Scott Erskine[1]  Janina Büscher[3]  Jürgen Titschack[4,5]  J Murray Roberts[2]  Sebastian Hennige[2]

[1]*School of Engineering and Physical Sciences, Institute of Mechanical, Process and Energy Engineering, Heriot-Watt University, Edinburgh, United Kingdom*
[2]*Changing Oceans Research Group, School of GeoSciences, University of Edinburgh, Edinburgh, United Kingdom*
[3]*GEOMAR Helmholtz Centre for Ocean Research Kiel, Biological Oceanography Research Group, Kiel, Germany*
[4]*Marum Center for Marine Sciences, University of Bremen, Bremen, Germany*
[5]*Senckenberg am Meer, Marine Research Department, Wilhelmshaven, Germany*
[*]*Corresponding author email: u.wolfram@hw.ac.uk*

**Abstract**

Ocean acidification is a threat to deep-sea corals and could lead to dramatic and rapid loss of the reef framework habitat they build. Weakening of structurally critical parts of the coral reef framework can lead to physical habitat collapse on an ecosystem scale, reducing the potential for biodiversity support. The mechanism underpinning crumbling and collapse of corals can be described via a combination of laboratory-scale experiments and mathematical and computational models. We synthesise data from electron back-scatter diffraction, micro-computed tomography, and micromechanical experiments, supplemented by molecular dynamics and continuum micromechanics simulations to predict failure of coral structures under increasing porosity and dissolution. Results reveal remarkable mechanical properties of the building material of cold-water coral skeletons of 462 MPa compressive strength and 45-67 GPa stiffness. This is 10 times stronger than concrete, twice as strong as ultrahigh performance fibre reinforced concrete, or nacre. Contrary to what would be expected, CWCs retain the strength of their skeletal building material despite a loss of its stiffness even when synthesised under future oceanic conditions. As this is on the material length-scale, it is independent of increasing porosity from exposure to corrosive water or bioerosion. Our models then illustrate how small increases in porosity lead to significantly increased risk of crumbling coral habitat. This new understanding, combined with projections of how seawater chemistry will change over the coming decades, will help support future conservation and management efforts of these vulnerable marine ecosystems by identifying which ecosystems are at risk and when they will be at risk, allowing assessment of the impact upon associated biodiversity.

# 1 Introduction

Ocean acidification is of concern to both tropical and cold-water coral (CWC) reefs [1]. It can cause a reduction in the growth rate of *live* coral [2], and dissolution of *dead* coral material (skeletons no longer covered in soft tissue) [3]. In tropical reefs, this can lead to a reduction in net growth rate [2, 4, 5]. For CWC reefs, found between 40 to 3,000 m deep [6], the threat is more significant and could result in dramatic and rapid habitat loss because of two factors: (i) CWC reef habitats and the biodiversity provision afforded by them is mostly provided by *dead* coral material. (ii) The aragonite saturation horizon (ASH), which is the depth at which aragonite (the calcium carbonate polymorph used by scleractinian corals to build their skeleton) becomes undersaturated, will rise above the majority (~70%) of CWC reefs over this century due to ocean acidification [1, 7]. Currently, most CWC reefs are above the ASH (aragonite concentration $\Omega_{Arag} > 1$) [8], and the few reefs found below the ASH ($\Omega_{Arag} < 1$, Figure 1a) have a marked absence of *dead* coral and low habitat complexity [9].

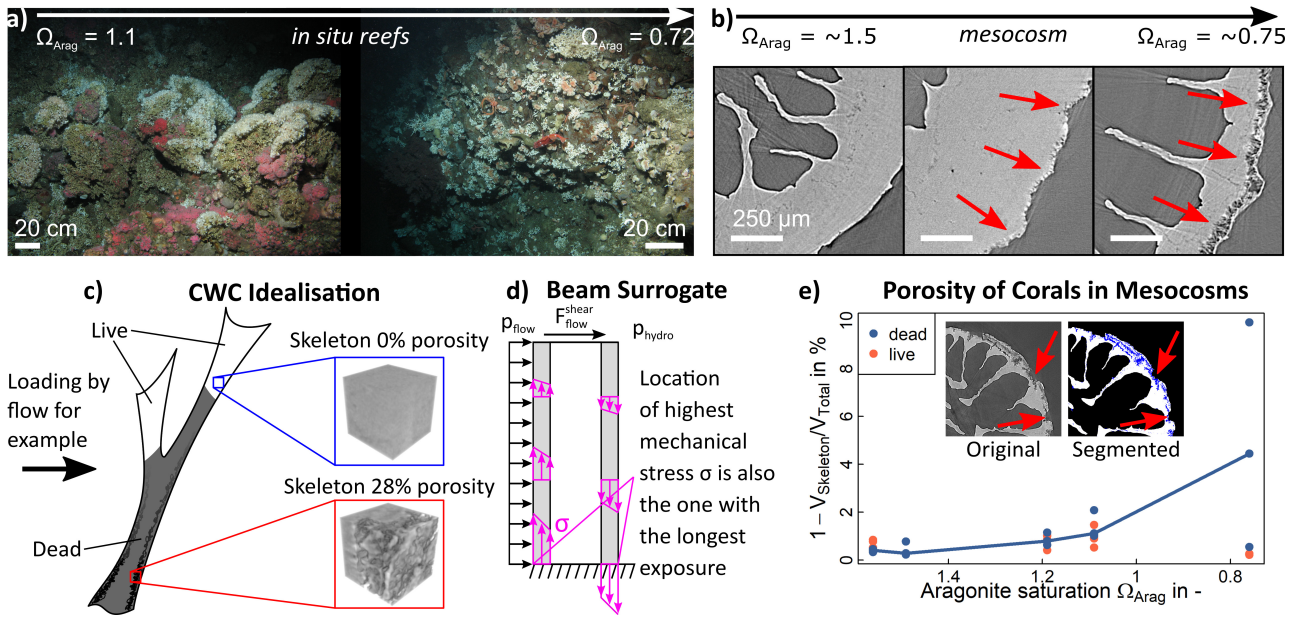

**Figure 1 Cold-water corals (CWCs) in an acidifying ocean: (a)** The transition of a high complexity reef with *live* coral (white tips) and *dead* coral (grey thicket) skeleton at $\Omega_{Arag} > 1$, to one with low complexity and no *dead* skeleton at $\Omega_{Arag} < 1$, is governed by collapse of the foundation framework due to rapidly increasing porosity in *dead* coral skeleton and subsequent dissolution when subjected to corrosive conditions (decreasing $\Omega_{Arag}$). **(b)** CWC skeletons exposed to future oceanic conditions for 12 months *in vitro* showed significantly increased porosity [9] (red arrows) for low $\Omega_{Arag}$. This was measured as skeleton pore volume (blue inlays in **(e)**) over total skeletal volume (white+blue inlays in **(e)**), which is a conservative approach. For a pre-defined volume of interest, e.g. the red one in **(c)** with 120 µm edge length, porosity was 28%. If the coral is simplified as a hollow beam loaded by, e.g. loads due to water flow ($p_{flow}, F_{flow}^{shear}$) and hydrostatic pressure ($p_{hydro}$), it becomes clear that porosity and dissolution affect the reef at its weakest point **(c-d)**. **(e)** Porosity increased significantly with decreasing $\Omega_{Arag}$ in *dead* but not in *live* skeletons [9]. (Images and data from [9, 10])

The occurrence of *live* coral below the ASH (Figure 1a) complements experimental research on *Lophelia pertusa* (also known as *Desmophyllum pertusum*) to date, indicating that *live* coral can continue to calcify under projected future temperature and ocean acidification conditions [11-16]. However, this ignores the larger and more serious ecosystem-scale threat to these habitats and associated biodiversity from a shoaling ASH; the potentially rapid loss of structural integrity and habitat complexity. This is of particular concern as high biodiversity associated with CWC reefs is strongly related to their structural complexity [17-19]. The majority of a CWC habitat is typically *dead* coral [20], with the living coral concentrated at the outermost parts of the reef where the coral is able to access and capture passing prey [5, 21]. This *dead* framework enhances the metabolic activity of CWC reefs and supports resource retention and recycling in the deep sea [22]. Coral framework also captures mobile sediment, leading to substantial deep-water coral reef and coral carbonate mound development [6]. The ability of *dead* coral framework to support living colonies by sustaining external loads is therefore of paramount importance not only for CWC habitat complexity and its ability to support other species, but for reef and mound formation and function.

To understand how future coral habitats may degrade, we need to consider the structural stability of the coral itself. Figure 1b and 1e suggest that loss of material, facilitated through an increase in porosity and a decrease in wall thickness, is exponentially related to a decrease in aragonite saturation [9]. This produces a relationship between a geochemical marker caused by ocean acidification (aragonite concentration) and a property affecting structural stability of the coral. To make sense of this, we can consider dissolution in a theoretical coral (Figure 1c) modelled as a hollow beam fixed at its base with a load applied from the side, e.g. pressure and shear force due to water flow as well as hydrostatic pressure (Figure 1d). Loss of material as increasing porosity and decreasing wall thickness in critical points of the outer mineralised skeleton increases fragility of the whole structure at its weakest point. This leads to early onset of mechanical failure, crumbling and collapse of CWCs.

The mechanism underpinning this shift can be described using mathematical and computational modelling that captures the complex structure of corals, combined with an appropriate material model that describes the mechanical stress-strain behaviour under load and, most importantly, that captures the impact of changing aragonite concentration. This allows prediction of failure rates of coral structures which, if combined with projections of how aragonite saturation will change over the coming decades, would allow us to estimate the timescales of this failure. Such an estimation would support future conservation and management efforts of these vulnerable marine ecosystems by understanding *which* ecosystems are at risk, *when* they will be at risk, and *how* much of an impact this will have upon associated biodiversity. However, understanding the multiscale material behaviour of CWCs and the mathematical and computational models that explain it are key gaps at present.

We aim to investigate the multiscale mechanical behaviour of CWC skeletons and analyse its deterioration with decreasing aragonite concentration and we develop a predictive model that can be interrogated for *in silico* experimentation. Specifically, our objectives are to (i) develop a multiscale material model that allows us to analyse multiscale mechanical consequences of ocean acidification for CWC skeletons; (ii) use dissolution data from long term experiments and *in situ* samples to analyse the impact of ocean acidification on the mechanical behaviour of the CWC skeleton; and (iii) demonstrate climate change induced increase in CWC fragility on a representative *L. pertusa* sample.

## 2 Materials and Methods

### 2.1 Coral samples

Coral samples examined here were previously collected [9, 16] and are from a gradient of environmental conditions ranging from $\Omega_{Arag} < 1$ to $\Omega_{Arag} > 1$ as detailed in Hennige et al. [9]. *In situ* samples from above and below the ASH, i.e. from acidified waters, were collected in the California Sea Bight $(\Omega_{Arag} = [0.71, 1.04])$. *In situ* samples from non-acidified waters $(\Omega_{Arag} = [1.67, 2.62])$ were collected from the Mingulay Reef Complex and Porcupine Seabight. All samples were covered with soft tissue prior to *in vitro* testing and are, therefore, considered to be *live* CWC skeletons.

### 2.2 CWC skeletons as multiscale, polycrystalline materials

CWCs form skeletons ranging across multiple length scales from micron sized crystals to millimetre sized wall structures that are affected by ocean acidification [9]. We propose to model the skeleton exposed to acidified waters as a three-scale, linear elastic material made up of needle shaped, crystalline building blocks that form a polycrystalline matrix (Figure 2).

CWC skeletons are made up of aragonite needles in the order of ≲5 µm long, 0.1-5wt% organic matrix embedded within the 3D framework [9, 23-26] as a thin film, and a nanometre sized porosity specified by Falini et al. [26] to be 3.9%. Aragonite needles protrude from rapid accretion deposits (RADs) [27] and, according to some sources, cluster into sclerodermites of diameter 10 µm x 50 µm [23]. The occurrence of sclerodermites is debated [23, 28] and we disregard this potential morphological feature as it is not important for our consideration. Needle accretion leads to CWC skeletons with wall thicknesses ≫ 100 µm, including translucent and opaque thickening bands [23, 24]. This composition suggests a polycrystalline setup where crystal needles are over an order of magnitude smaller than the structural volume they create (Figure 2). Previous studies reported a certain degree of ordering that is generated by thickening growth of the skeleton [28-30], and ordered crystal arrangements would introduce a certain direction dependence (anisotropy) of the mechanical properties at the skeletal wall level. To assess this, we reanalysed previously published [9, 16] scanning electron microscopy (SEM) and electron backscatter diffraction (EBSD) data of 10 *L. pertusa* samples (Supplementary materials

Figure S1) with regards to crystal orientation. A random crystal assembly was observed in SEM surface images (Figure S1). EBSD images also show that 38.5% crystals are oriented in (001), 27.6% in (100), and 33.9% in (010) direction across their cross-sectional areas which are also varying. This is consistent with observations by Pasquini et al. [31] who reported no preferential organisation of the aragonite needles based on combined data from nanoindentation, electron microscopy, atomic force microscopy, and X-ray diffraction patterns of longitudinal and transverse sections in solitary *Balanophyllia europaea* and colonial *Stylophora pistillata* corals. Crystals protruding from RADs [27] in spheroidal fashion can produce such isotropic behaviour and appear as observed in our SEM and EBSD images. Mechanically, this can be considered similar to a random aragonite crystal assembly.

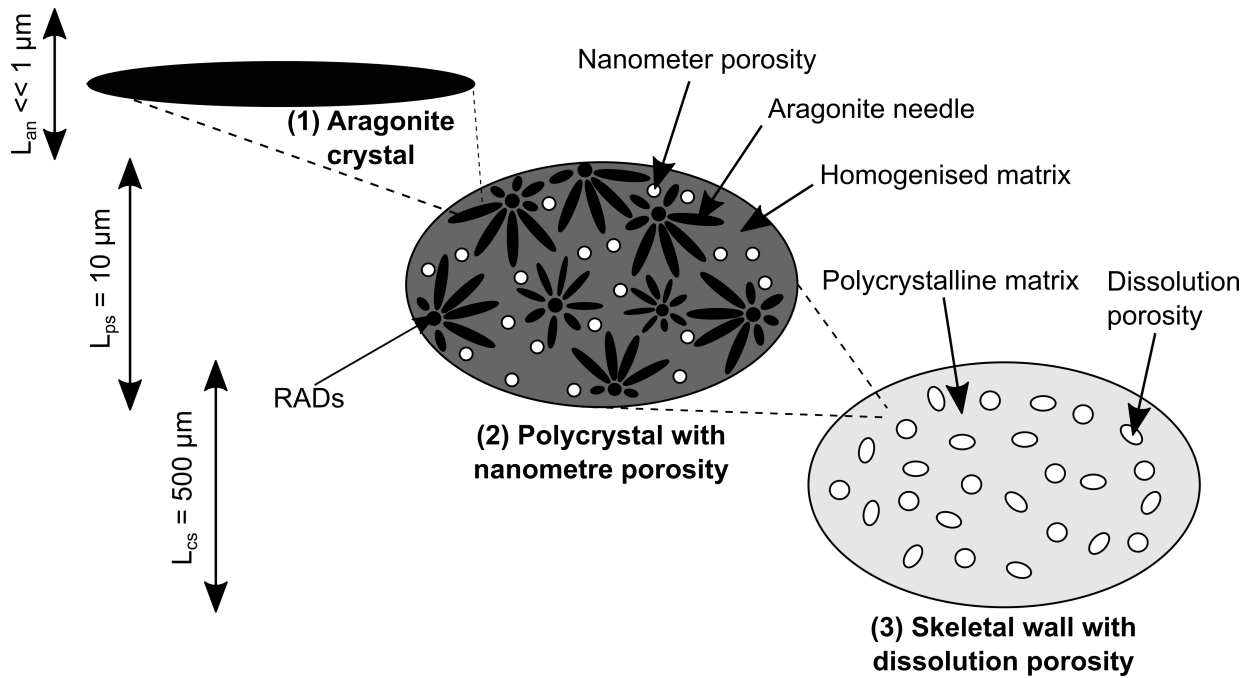

**Figure 2 Multiscale model of the skeletal wall:** The skeleton of *L. pertusa* consists of needle shaped aragonite crystals with finite aspect ratio. These needles are arranged around rapid accretion deposits and form a polycrystalline matrix which then forms the coral wall. This wall is subject to ocean acidification which generates spheroidal pores.

Aragonite needles typically show aspect ratio ranges of 1-100, with the most common aspect ratio to be in the range of 8-12 [32-35]. To corroborate this in our samples, we analysed inverted pole figure images of our EBSD data (Figure S1) to determine crystal aspect ratios. As crystals grow along the (001) axis, we randomly picked 10 crystal needles in 10 samples displaying the (010) or (100) axes and measured the longest vs. shortest axis using ImageJ (V1.52p, National Institutes of Health). The median aspect ratio found in the EBSD cross-sections was 4.86 (2.01-13.85). As these cross-sectional areas represent conic cuts through the crystals (Figure S1), we infer that the lower range of our measurements are distorted but that the upper range represents a good approximation for aragonite needles. These values are comparable to values from the literature [32-35] and justify approximating aragonite crystals with a prolate spheroidal shape.

These considerations allow us to use continuum micromechanics [36] to determine the mechanical properties and study the impact of ocean acidification on the mechanical behaviour on the relevant length scales of the CWC skeleton [9]. In this framework, the exact crystal shape is not critical as we can capture their influence on the mechanical properties through their volume fraction, aspect ratio [37, 38], and an appropriate representation of the crystal inclusion [39, 40].

**2.3  Dissolution, porosity, and affected layer**

Hennige et al. [9] observed acidification induced porosity using synchrotron radiation computed micro-tomography (SRµCT) in samples from the Southern California Bight, a location where CWCs live in conditions which are analogous to those most CWC reefs may experience by 2099. This has also been observed in Mediterranean coral species subjected to ocean acidification conditions [41, 42]. Although not directly comparable to CWCs due to the zooxanthellate nature of these corals and the higher aragonite concentration at the volcanic test sites, the occurrence of porosity due to acidified waters supports the dissolution theme and illustrates applicability of our modelling beyond CWCs. Hennige et al. [9] corroborated *in situ* findings with SRµCT analyses of samples from 12 months *in vitro* experiments that mimicked future oceanic conditions (Figure 1b). When analysing their samples, Hennige et al. [9] used a conservative measure and related total pore volume to total volume of the coral sample which resulted in low porosity values (Figure 1e) with porosity measured locally being considerably higher. We reanalysed eight samples treated in *in vitro* mesocosm experiments [9] (Figure 1b) to extract an affected layer thickness and a maximum porosity per volume of interest.

Porosity was quantified similar to Hennige et al. [9] (Figure S2) and the affected layer thickness was determined using the alpha shape toolbox in Python (Python 3.7). A shape parameter of $\alpha = 0.05$ was selected and used in a slice-by-slice extraction of the affected layer (Figure S2). Radial thickness of this layer was calculated by first determining the medial axis of the shape (Python 3.7, Scikit-image, morphology.medial_axis) and then performing a distance transform. We extracted pores using a connected component analyses (Python 3.7, Scikit-image: measure.label) and measured their degree of anisotropy (Python 3.7 PoreSpy, metrics.regionprops_3D) as the major axis length over the minor axis length of a pore.

**2.4  Aragonite single crystal elasticity and strength**

To develop the envisioned material model, stiffness and strength of the constituent crystal phase are needed. While single crystal stiffness is available from the literature (Table 1), single crystal strength needs to be determined as it may serve as an upper bound of the strength of the polycrystal. CWC skeletons fail in a brittle fashion [16] involving crystal decohesion, which is an interfacial phenomenon, rather than breakage of individual crystals. We, therefore, consider interfacial decohesion as a lower bound of the strength of the polycrystal. For gypsum [38] and bone [43] it has been proposed that interfacial decohesion of crystal needles are governed by tensile and shear stresses associated with the needle direction.

**Table 1 Stiffness values for aragonite:** Experimental and computational stiffnesses ($C_{ij}$) for aragonite crystals from the literature as well as stiffnesses computed in this study.

| Source | $C_{11}$ | $C_{22}$ | $C_{33}$ | $C_{44}$ | $C_{55}$ | $C_{66}$ | $C_{23}$ | $C_{31}$ | $C_{12}$ |
|---|---|---|---|---|---|---|---|---|---|
| **Experimental values in GPa** | | | | | | | | | |
| Liu et al. [44] | 171.1 | 110.1 | 98.4 | 39.3 | 24.2 | 40.2 | 41.9 | 27.8 | 60.3 |
| Voigt [45] | 159.6 | 87.0 | 85.0 | 41.3 | 25.6 | 42.7 | 15.9 | 2.0 | 36.6 |
| Hearmon [46] | 228.8 | 124.2 | 85.0 | 82.6 | 51.2 | 94.4 | 18.9 | 10.6 | 102.4 |
| Min | 159.6 | 87.0 | 85.0 | 39.3 | 24.2 | 40.2 | 15.9 | 2.0 | 36.6 |
| Median | 171.1 | 110.1 | 85.0 | 41.3 | 25.6 | 42.7 | 18.9 | 10.6 | 60.3 |
| Max | 228.8 | 124.2 | 98.4 | 82.6 | 51.2 | 94.4 | 41.9 | 27.8 | 102.4 |
| **Computational values in GPa** | | | | | | | | | |
| Fisler et al. [47] | 155.3 | 104.2 | 89.9 | 36.7 | 12.4 | 23.3 | 48.0 | 54.7 | 55.9 |
| Xiao et al. [48] | 174.8 | 112.9 | 104.7 | 40.1 | 26.6 | 45.6 | 56.1 | 41.1 | 67.9 |
|  | 164.4 | 112.0 | 59.2 | 40.5 | 33.9 | 49.0 | 48.2 | 39.0 | 65.3 |
| Pavese et al. [49] | 157.7 | 100.9 | 68.3 | 36.0 | 24.1 | 41.1 | 50.4 | 34.1 | 58.0 |
|  | 194.2 | 117.1 | 71.3 | 44.1 | 34.5 | 43.8 | 50.2 | 35.7 | 65.9 |
| This study | 156.1 | 96.0 | 66.5 | 34.1 | 24.6 | 38.7 | 50.8 | 36.1 | 59.1 |
| Min | 155.3 | 96.0 | 59.2 | 34.1 | 12.4 | 23.3 | 48.0 | 34.1 | 55.9 |
| Median | 161.1 | 108.1 | 69.8 | 38.4 | 25.6 | 42.5 | 50.3 | 37.6 | 62.2 |
| Max | 194.2 | 117.1 | 104.7 | 44.1 | 34.5 | 49.0 | 56.1 | 54.7 | 67.9 |

We computed single crystal strength using molecular dynamics simulations assuming that twinned crystal boundaries are of similar strength than the single crystal. The simulations were carried out the using open-source software LAMMPS (Large Scale Atomic/Molecular Massively Simulator, https://lammps.sandia.gov) developed by Plimpton [50]. The unit cell dimensions of aragonite were a=4.961, b=7.967 and c=5.740 Å with 4 calcium, 4 carbon and 12 oxygen atoms. By following Pavese et al. [49], the Born-type potential was utilised to account for interactions between Ca-O d C-O and O-O. The dispersive term was included only for O-O interactions. The covalent bond in the $CO_3$ molecule was modelled by harmonic angular and torsional potentials. The potential parameters were taken from the RIM2 model of Pavese et al. [49] (see their Table 4 for details). Before performing our tests, we equilibrated atoms by performing Nosé-Hoover thermostat sampled from isothermal-isobaric ensemble (NpT) at 298 K and 1 bar for 48 ps. For uniaxial tensile tests, the flow configuration was a rectangular box with an aspect ratio of 4 containing 24-unit cell along the tensile direction. For shear tests, we used a cubic box with 13824 unit cells (24 cells in each direction). For all simulations, we employed periodic boundary conditions in all directions and the time-step was set to 0.001 ps. We applied the explicit deformation method, where the dimension of box was changed at a constant engineering strain rate ($\varepsilon = 0.01$ 1/s unless specified otherwise) to compute elastics constants. The elements of the stiffness tensor were extracted from three tension and three shear deformations following Clavier et al. [51]. Strength of the crystal was measured as the maximum stress for each loading mode (Figure 3, Table 2).

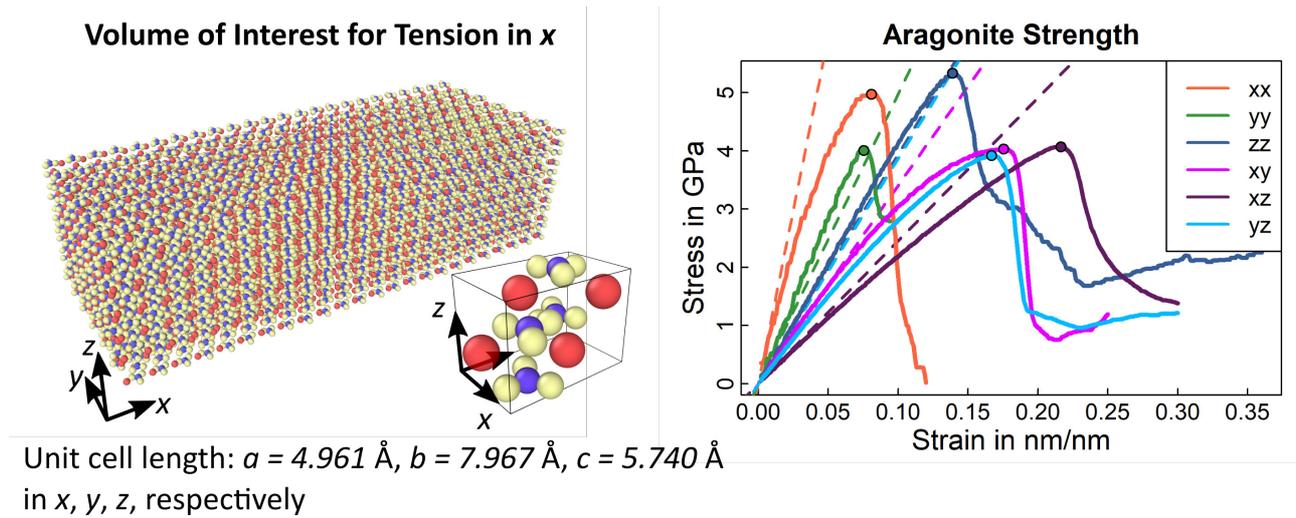

Unit cell length: $a$ = 4.961 Å, $b$ = 7.967 Å, $c$ = 5.740 Å in $x$, $y$, $z$, respectively

**Figure 3 Aragonite stiffness and strength:** An aragonite volume of interest with 24x6x6 unit cells of dimension $a \times b \times c$ was used to determine crystal strength. The left image shows the setup used in the tensile tests along x and an animation of this test is shown in Video S1 and S2 (supplementary material). The model was verified against experimental stiffness values (Table 1) before conducting the strength test whose results are shown on the right. Maximum bearable stress is indicated by solid circles in each stress-strain curve in the image on the right.

Our molecular dynamics analyses provided results for a perfect, theoretical crystal. To propose a failure criterion, we need to identify tensile and shear stresses of the interface between two crystals, compare them to these single crystal results, and use the lower set of values to model failure at the crystal length scale (Section 2.6). To our knowledge, interfacial tensile and shear stresses for CWC skeletal aragonite needle assemblies have not been measured and we consider nacre as a proxy-material (Table 2). Like CWC skeletons, nacre is made of aragonite but in the form of platelets that are impregnated with a protein layer. It has volume fractions of 95wt% aragonite and 5wt% organic material [52] that are comparable to CWC skeletons. Nacre platelets loaded along the in-plane axis, as for example during compression bending of the shell, resemble a similar loading situation as the aragonite needles in our case with tensile and shear stresses associated with the platelet axis governing the failure. Barthelat et al. [52] argued that dry nacre approximates the brittle, interfacial yield strength of pure aragonite which resembles the situation in our material model (Figure 2).

**Table 2 Aragonite and polycrystal yield strength:** Yield strengths for the aragonite needle and the polycrystalline coral skeleton not exposed to acidified waters. $e_i$ with $i = (x, y, z)$ denote directions of testing in the molecular dynamics test. $\sigma_{int}^{ut}$ and $\sigma_{int}^{us}$ denote interfacial tensile and shear strength of the aragonite crystal while $\sigma_{poly}^{UT}$ and $\sigma_{poly}^{UC}$ are the tensile and compressive strengths of the polycrystalline skeleton not exposed to acidified waters, respectively.

| Aragonite needle | $e_x \otimes e_x$ | $e_y \otimes e_y$ | $e_z \otimes e_z$ | $e_y \otimes e_z$ | $e_z \otimes e_x$ | $e_x \otimes e_y$ |
|---|---|---|---|---|---|---|
| $\sigma_{Arag}$ in GPa | 4.97 | 4.00 | 5.33 | 4.03 | 4.06 | 3.92 |
| Average in GPa | | 4.77 | | | 4.00 | |

| Aragonite interfacial* | $\sigma_{int}^{ut}$ in MPa | $\sigma_{int}^{us}$ in MPa |
|---|---|---|
| Nacre interfacial strength | 158.33 | 70.00 |
| Based on micropillar tests | 294.49 | 130.20 |

*Barthelat et al. [52] argued that dry nacre approximates the brittle, interfacial yield strength of pure aragonite. We assume that a nacre platelet loaded along the in-plane axis resembles a similar loading situation as the aragonite needles in the CWC skeleton. Consequently, we use tensile [52-54] and shear [52] strengths for nacre as a lower bound, with a review of tests given by Sun and Bhushan [55].

| Polycrystal at $\phi_{np} = 3.9\%$ | $\sigma_{poly}^{UT}$ in MPa | $\sigma_{poly}^{UC}$ in MPa |
|---|---|---|
| Nacre interfacial | 95.45 | 248.52 |
| Based on micropillar tests | 177.53 | 462.25 |

## 2.5 Elasticity of the CWC skeleton not exposed to acidified waters

We enforced transverse isotropy of the median experimental stiffness tensor of aragonite (Table 1) to reduce the computational costs of determining skeletal stiffness and strength, so that $C_{00} = \frac{1}{2}(C_{11} - C_{22})$, $C_{03} = \frac{1}{2}(C_{31} - C_{23})$, $C_{01} = C_{12}$, and $C_{66} = \frac{1}{2}(C_{11} - C_{12})$. This simplification results in slightly lower values than the experimental median but is within the range of values reported for aragonite (Table 1) as well as calcite [56], another polymorph of calcium carbonate. We modelled the coral skeleton (CS) as a polycrystal consisting of randomly oriented aragonite needles with stiffness $\mathbb{S}_{Arag}$ and spherical nano-porosity ($\phi_{np}$) of 3.9% [26] as reference value. This random needle setup warranted a self-consistent scheme [40] which we modified to include transverse isotropic crystals following proposals for gypsum and bone by Sanahuja et al. [38] and Fritsch et al. [57], respectively. We formulated stiffness of the polycrystalline CWC skeleton based on the median experimental aragonite stiffness (Table 1). In addition to $\mathbb{S}_{Arag}$, the effective polycrystalline stiffness tensor, $\mathbb{S}_{CS}$, depends on the volume fraction of the crystals $(1 - \phi_{np})$, their aspect ratio, as well as $\phi_{np}$ that is filled with organic matter which we assumed does not contribute to the stiffness of the skeletal wall. The homogenised stiffness tensor for the coral skeleton can then be given as:

$$\mathbb{S}_{CS} = (1-\phi_{np}) \int_{\varphi=0}^{2\pi} \int_{\theta=0}^{\pi} \mathbb{S}_{Arag}(\varphi,\theta) : \left(\mathbb{I} + \mathbb{P}_{ndl}^{CS}(\varphi,\theta)\left(\mathbb{S}_{Arag}(\varphi,\theta) - \mathbb{S}_{CS}\right)\right)^{-1} \frac{\sin\theta}{4\pi} d\theta d\varphi :$$

$$\left((1-\phi_{np}) \int_{\varphi=0}^{2\pi} \int_{\theta=0}^{\pi} \left(\mathbb{I} + \mathbb{P}_{ndl}^{CS}(\varphi,\theta)\left(\mathbb{S}_{Arag}(\varphi,\theta) - \mathbb{S}_{CS}\right)\right)^{-1} \frac{\sin\theta}{4\pi} d\theta d\varphi + \phi_{np}\left(\mathbb{I} + \mathbb{P}_{sph}^{CS}\left(\mathbb{S}_{np} - \mathbb{S}_{CS}\right)\right)^{-1}\right)^{-1}. \quad (1)$$

Therein, $\mathbb{I}$ is the 4$^{th}$ order identity tensor ($\mathbb{I}_{ijkl} = (\delta_{ik}\delta_{jl} + \delta_{il}\delta_{jk})/2$), $\mathbb{P}_{Arag}^{CS}$ is the 4$^{th}$ order Hill tensor which is defined by $\mathbb{P}_{ndl}^{CS} = \mathbb{R} : \mathbb{S}_{CS}^{-1}$ and $\mathbb{R}$ is the 4$^{th}$ order Eshelby tensor for the chosen inclusion problem. The Eshelby tensor solutions for inclusion shapes [39] used in this study are provided in supplementary material Section S3. The ':' symbol denotes the double inner product of two tensors. $\mathbb{S}_{CS}$ is the effective tensor stiffness of the coral skeleton whose determination requires an iterative solution approach (Mathematica 12.1).

## 2.6 Strength of the CWC skeleton not exposed to acidified waters

The failure mechanism for CWC skeletons is brittle with decohesion and breakage of crystal arrangements [16]. We modelled failure of the most adversely loaded aragonite crystal using a Mohr-Coulomb criterion [38, 58]

$$\sigma_{nn} + \beta\sigma_{nt} = \sigma_{int}^{ut} \quad (2)$$

with $\sigma_{nn}$ the stress in normal direction, $\sigma_{nt}$ the stress in tangential direction and $\beta = \sigma_{int}^{ut}/\sigma_{int}^{sh}$ the ratio between tensile and shear yield stress of the aragonite crystal (Table 2, Figure 4). We switched length scales between the polycrystal arrangement and the aragonite needle using concentration tensor $\mathbb{B}(\varphi,\theta)$ [38, 57, 59, 60]

$$\mathbb{B}(\varphi,\theta) = \mathbb{S}_{Arag}(\varphi,\theta) : \left[\left(\mathbb{I} + \mathbb{P}_{ndl}^{CS}(\varphi,\theta):\left(\mathbb{S}_{Arag}(\varphi,\theta) - \mathbb{S}_{CS}\right)\right)^{-1} : \right.$$

$$\left.\left((1-\phi) \int_{\varphi=0}^{2\pi}\int_{\theta=0}^{\pi} \left(\mathbb{I} + \mathbb{P}_{ndl}^{CS}(\varphi,\theta)\left(\mathbb{S}_{Arag}(\varphi,\theta) - \mathbb{S}_{CS}\right)\right)^{-1} \frac{\sin\theta}{4\pi} d\theta d\varphi + \phi\left(\mathbb{I} + \mathbb{P}_{sph}^{CS}\left(\mathbb{S}_{np} - \mathbb{S}_{CS}\right)\right)^{-1}\right)^{-1}\right] : \mathbb{S}_{CS}^{-1} \quad (3)$$

so that, the average stress in an aragonite needle with direction $\boldsymbol{n} = \boldsymbol{n}(\varphi, \vartheta)$ (Figure S3) in the polycrystal due to a macroscopic stress on the polycrystal becomes:

$$\boldsymbol{\sigma}_{Arag}(\varphi,\theta) = \mathbb{B}(\varphi,\theta) : \boldsymbol{S} = \mathbb{B}(\varphi,\theta) : S\,\widehat{\boldsymbol{S}}. \quad (4)$$

In this equation $S = ||S||$ is the magnitude of the stress tensor and $\hat{S} = S/||S||$ its direction. We computed the stress components along the needle long axis $\boldsymbol{n} = \boldsymbol{n}(\varphi, \theta)$ and the one tangential to the needle long axis $\boldsymbol{t} = \boldsymbol{t}(\varphi, \theta, \psi)$ (Section S4) as

$$\sigma_{nn} = \boldsymbol{n} \cdot \boldsymbol{\sigma}_{Arag}(\varphi, \theta) \cdot \boldsymbol{n} \quad \text{and} \quad \sigma_{nt} = \boldsymbol{n} \cdot \boldsymbol{\sigma}_{Arag}(\varphi, \theta) \cdot \boldsymbol{t}$$

$$\sigma_{nn} = \boldsymbol{n} \cdot (\mathbb{B}(\varphi, \theta) : S) \cdot \boldsymbol{n} \quad \text{and} \quad \sigma_{nt} = \boldsymbol{n} \cdot (\mathbb{B}(\varphi, \theta) : S) \cdot \boldsymbol{t}.$$

(5)

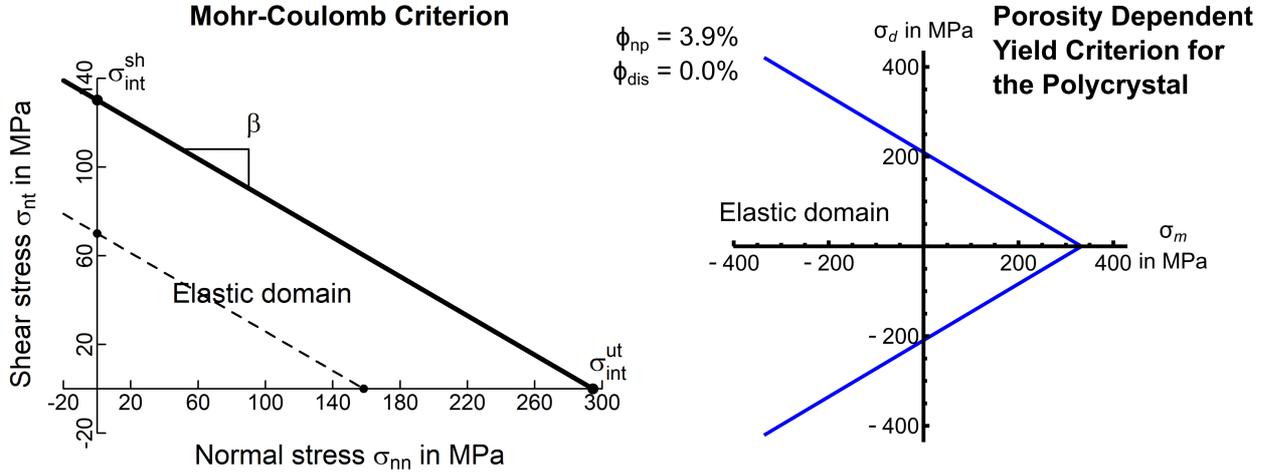

**Figure 4 Yield criteria for the cold-water coral skeleton:** Left image shows an interfacial Mohr-Coulomb criterion (equation 3) of the aragonite crystals using interfacial tensile $\sigma_{int}^{ut}$ and shear $\sigma_{int}^{sh}$ strengths based on micropillar tests (solid line) as well as for nacre aragonite interfacial strengths (dashed line). Right image shows a porosity dependent quadric yield criterion [61] plotted along its trisectrix with mean stress $\sigma_m$ and deviatoric stress $\sigma_d$. The criterion can incorporate increasing porosities (Figure 7) due to acidification and erosion processes and is plotted here with no dissolution porosity $\phi_{dis}$. Since its material parameters are derived from the underlying polycrystal elasticity and crystalline yield behaviour, its parameters contain a $\phi_{np} = 3.9\%$.

We used equation (2) and the interfacial strengths of aragonite (Table 2) to calculate strength for the polycrystalline skeletal wall for different stress directions $\hat{S}$

$$\text{Max} \quad \boldsymbol{n} \cdot (\mathbb{B}(\varphi, \theta) : \hat{S}) \cdot \boldsymbol{n} + \beta |\boldsymbol{n} \cdot (\mathbb{B}(\varphi, \theta) : \hat{S}) \cdot \boldsymbol{t}|$$

$$\text{with } 0 = \varphi, 0 \leq \theta < \pi, 0 \leq \psi < 2\pi$$

(6)

This yielded the direction of maximum stress. We discretised the angular range with increments of $\Delta\theta = \pi/128$ and $\Delta\psi = \pi/32$ with the aim to find the maximum of equation (6) (NMaximize, Mathematica 12.1) and, thus, failure associated with the most adversely loaded crystal [62]. Evaluating equation (2) for these maximal stress directions allowed us to calculate corresponding, micromechanically informed, elastic limits in tension $\sigma_{poly}^{ut}$ and compression $\sigma_{poly}^{uc}$ for the polycrystalline skeleton that is not affected by ocean acidification, but which incorporates $\phi_{np} = 3.9\%$:

$$S = \frac{\sigma_{Int}^{ut}}{\boldsymbol{n} \cdot (\mathbb{B}(\varphi, \theta) : \hat{S}_{33}) \cdot \boldsymbol{n} + \beta |\boldsymbol{n} \cdot (\mathbb{B}(\varphi, \theta) : \hat{S}_{33}) \cdot \boldsymbol{t}|}.$$

(7)

We considered compression and tension along $e_3 = (0,0,1)$ to calculate the yield strengths of the polycrystal in tension $\sigma_{poly}^{ut}$ and compression $\sigma_{poly}^{uc}$ following an approach used for quantifying yield envelops for other mineralised tissues [63]. Compression and tension are sufficient to identify a polycrystalline yield envelope as we assumed isotropy (Section 2.2). The resulting polycrystalline yield criterion (Figure 4) features a conical shape (Drucker-Prager failure criterion) with tensile and compressive strengths $\sigma_{poly}^{ut}$ and $\sigma_{poly}^{uc}$ based on $\sigma_{int}^{ut}$ and $\sigma_{int}^{sh}$ (Table 2). The conical shape has the additional benefit that it reflects adaption to the large hydrostatic pressures encountered by CWCs that suggest that incompressibility under triaxial loading is important.

### 2.7 Micropillar compression tests of the skeleton not exposed to acidified waters

To complement our predictive modelling with experimental strength tests, we performed micropillar compression tests adopting previous protocols [64, 65] on $n = 12$ CWC samples (Section 2.1, Section S8). We selected samples that were covered with soft tissue upon retrieval. These can, thus, be considered *live* coral skeletons, not affected by ocean acidification [9]. We selected three samples from the Mingulay Reef Complex and three from the Porcupine Seabight in the NE Atlantic that represent non-corrosive oceanic conditions as well as six samples from the California Sea Bight which were collected at or below the ASH which represents a corrosive oceanic environment [9].

We repolished the samples using previous preparation protocols [9] and fabricated micropillars using ultrashort pulsed laser ablation (Section S5). On each of the 12 CWC samples two arrays of six micropillars were manufactured (Figure S4). For each sample, we selected six of the micropillars and conducted destructive micropillar compression tests using a custom-made portable microindenter (Alemnis AG, Switzerland) equipped with an 88 µm diameter conical diamond punch [64, 65]. We calibrated frame compliance prior to testing using a Berkovich probe and a standard fused quartz sample [66]. Each pillar was compressed uniaxially at a rate of 0.05 µm/s to a total displacement of 10.25 µm. Partial unloading was conducted by retracting the probe 0.25 µm for every 0.75 µm the probe travelled. Load and displacement were recorded simultaneously throughout each test at 30 Hz.

The large taper angle that results from using laser ablation (Figure S4) without subsequent focussed ion beam milling [64, 65] did not allow direct conversion of experimentally observed forces into stresses. To interpret the micropillar compression tests we implemented our material model into an elasto-viscoplastic framework (Sections S7, S7.1) as a user defined material for Abaqus (v6.16, Dassault Systémes). We generated a finite element model of the micropillar (Section S7, Figure S6) and conducted *in silico* micropillar compression tests similar to our experimental ones.

## 2.8 Incorporating the impact of ocean acidification on skeletal stiffness and strength

After creating a model for the skeleton not affected by ocean acidification, we incorporate the mechanical impact of ocean acidification on stiffness and strength. We determined a significant increase of porosity after subjecting CWC samples to acidified water resembling a future ocean (Figure 1c and e, Figure S2). Assuming the shape of these pores to be spheroidal with random orientations allowed us to model dissolution as spherical inclusion. As we obtained an isotropic stiffness tensor for the mineralised coral skeleton (Section 2.5), we used a Mori-Tanaka scheme [59, 67] to incorporate acidification induced porosity (Figures 1, 2, Figure S2). Both the coral skeleton (matrix) and the pores (inclusion) can be represented by their respective stiffness tensors ($\mathbb{S}_{CS}$ and $\mathbb{S}_{PO}$, respectively) and volume fractions $\phi$, such that $\phi_{CS} + \phi_{OA} = 1$. The inclusions are assumed to be subject to the same homogeneous load so that the skeleton exposed to acidified waters represents a matrix-inclusion type composite [36] and its stiffness can be written as:

$$\mathbb{S}_{OA} = \left(\phi_{CS}\mathbb{S}_{CS} + \phi_{PO}\mathbb{S}_{PO} : \left(\mathbb{I} + \mathbb{P}_{PO}^{MT}(\mathbb{S}_{PO} - \mathbb{S}_{CS})\right)^{-1}\right) : \\ \left(\phi_{CS}\mathbb{I} + \phi_{PO}\left(\mathbb{I} + \mathbb{P}_{PO}^{MT}(\mathbb{S}_{PO} - \mathbb{S}_{CS})\right)^{-1}\right)^{-1}. \tag{8}$$

This stiffness tensor incorporates acidification induced porosity into the reversible mechanical behaviour of the CWC skeleton. We then used tensile and compressive strengths $\sigma_{poly}^{ut}$ and $\sigma_{poly}^{uc}$ of the polycrystalline CWC skeleton not exposed to acidified waters (equation 6, Section 2.6), to incorporate acidification induced porosity into a failure criterion of the CWC skeleton. Based on our considerations in Section 2.6, we used a conic criterion that was first proposed by Maghous et al. [68] and generalised by Schwiedrzik et al. [61]:

$$Y_{OA}(S) := \sqrt{S : \mathbb{F} S} + F : S - 1 = 0$$

$$\text{with} \quad \mathbb{F} = \frac{1 + \frac{2}{3}\phi_{OA}}{(1 - \phi_{OA})^2 h^2 T^2}(I \overline{\otimes} I) - \frac{\frac{1}{3} + \frac{1}{18}\phi_{OA}}{(1 - \phi_{OA})^2 h^2 T^2}(I \otimes I), \text{ and } F = \frac{1}{3(1 - \phi_{OA})h}I \tag{9}$$

in which cohesion $h = \frac{2}{3}\frac{\sigma_{poly}^{ut}\sigma_{poly}^{uc}}{\sigma_{poly}^{uc} - \sigma_{poly}^{ut}}$ and friction coefficient $T = \sqrt{6}\frac{\sigma_{poly}^{uc} - \sigma_{poly}^{ut}}{\sigma_{poly}^{uc} + \sigma_{poly}^{ut}}$.

$I$ denotes the identity tensor, $A \overline{\otimes} B = (A_{ik}B_{jl} + A_{il}B_{jk})/2)$ the symmetric product, and $\otimes$ the tensor product or dyad. Section S6 illustrates how cohesion $h$ and friction coefficient $T$ are derived and Section S7 details how this was implemented into an elasto-viscoplastic framework within the finite element package.

## 2.9 The impact of ocean acidification on an exemplary coral colony

To show the impact of ocean acidification on a real coral structure, we used the developed material model in an image based finite element model of a coral specimen of *L. pertusa* (Figure 5, Figure S6). The coral specimen was scanned in a clinical CT (Toshiba Aquilon 64) with an X-ray source voltage of 120 kV and a current of 600 mA [69]. The resulting CT image was reconstructed with a voxel size of 0.35 mm in plane, 0.3 mm slice distance, and exported to DICOM-format.

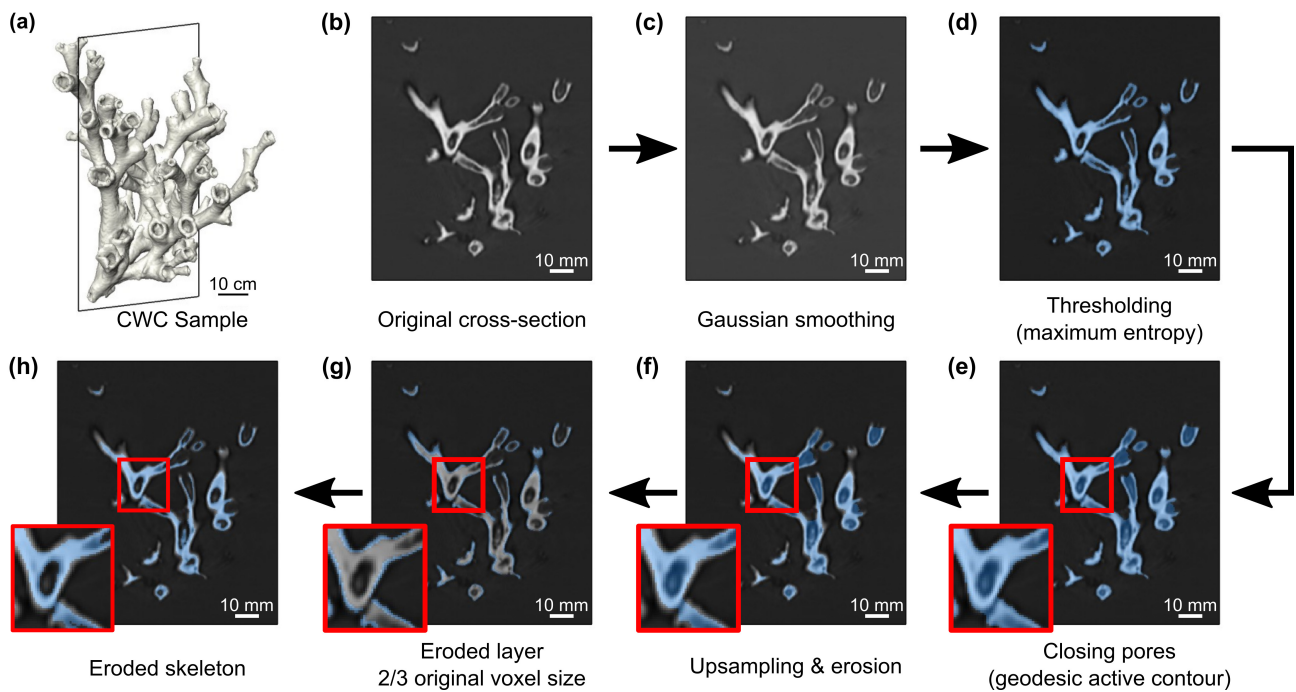

**Figure 5 Incorporating dissolution on a representative coral specimen: (a)** 3D render of the exemplary cold-water coral (CWC) specimen. **(b)** Original cross-section through the XZ plane; **(c)** Gaussian filter was applied ($\sigma = 0.3$); **(d)** Coral skeleton was segmented using Maximum Entropy algorithm threshold; **(e)** A 3D morphological geodesic active contour algorithm was implemented to detect the outer surface and fill the pores and cavities in the skeleton. **(f)** The contour image was upsampled and eroded in steps of 1/3 of the original voxel size. **(g)** 0.23 mm eroded layer (blue) superimposed to the original CT cross-section (grey) and 0.23 mm eroded layer (blue). **(h)** Final image composite of the eroded coral skeleton (blue) and the original CT cross-section (grey). See Video S3 for a visualisation of the image processing steps as well as Figure S6 for the finite element models.

We subjected the model to an increased porosity and acidification induced thinning. The impact of porosity was captured through our micromechanical material model. Thinning due to acidification, i.e. loss of wall thickness, was induced on the outer surface of the coral and was implemented in Python using *SimpleITK* and *Scikit-Image* libraries (Figure 5, Video S3). Images were first smoothed using a recursive Gaussian filter ($\sigma = 0.3$) and binarized using the Maximum Entropy algorithm. Unconnected regions were removed, and a 3D morphological geodesic active contour algorithm [70] was employed to detect the outer surface of the coral specimen and mask-out both the coral skeleton and internal pores and cavities. The resulting contour image was up-sampled by a factor of 3 in x- and y-direction and a morphological erosion of approximately 0.12 mm, 0.23 mm, or 0.35 mm (1/3, 2/3 and 3/3 of the original voxel size) was performed to simulate progressive thinning of the outer layer (Figure S2) and which was motivated by the thickness of the affected layer identified in Sections 2.3 and 3.1. Finally, the binarized image of the coral specimen was multiplied by the eroded contour images to obtain a final 3D binary dataset of the eroded coral specimen. The four resultant 3D images

(i.e. no thinning and three erosion steps) were converted to tetrahedral meshes with four-noded tetrahedral elements with maximum size of approximately 0.5 mm using *pygalmesh*. We then imported these meshes into Matlab (R2020a), assigned boundary conditions, and generated input files for Abaqus (Section S7.2, Figure S6). Boundary conditions mimicked a contact pressure load on the coral skeleton surface simulating an artificially chosen sea current with a velocity of 3 m/s in a direction perpendicular to the longitudinal axis of the skeleton (Section S7.2, Figure S6). This velocity was chosen to firmly overload the sample and represents three times the maximum water current reported by Haugan et al. [71] (0.3-1 m/s). Note that this is an academic example to illustrate the effect of increasing velocity and loss of skeletal wall thickness.

### 2.10 Statistical analyses

Statistical analyses were performed using Gnu R[72] (Rstudio 1.4.1103). To test normal distribution of data, quantile-quantile plots and Shapiro-Wilk post-hoc tests were used. If normality was given, groups were compared using Student's t-tests otherwise Wilcoxon rank sum tests were used. We assumed a significance level of $p = 0.05$. Sample data is presented by means of distribution independent median and ranges.

## 3 Results

### 3.1 Dissolution, porosity, and affected layer

Affected layer thickness was found to be up to 109.17-253.13 µm and the affected layer contained porosities of 10.6-33.3% (Figure S2). Using the connected component analyses we extracted 3135825 ± 2098230 pores/sample. The average degree of anisotropy was found to be 1.9 which suggests a spheroidal shape of dissolution pores. These pores were however not aligned but randomly oriented, which justifies approximation of the pore shape by a sphere with aspect ratio 1. Given this, the assumptions made in Section 2.8 for the pore shape are justified and usable to investigate the mechanical impact of ocean acidification on these CWC skeletons.

### 3.2 Aragonite single crystal elasticity and strength

Aragonite single crystal elasticity determined using molecular dynamics was in good agreement with experimental and computational values from the literature (Table 1). Therefore, we consider our models suitable to determine single crystal strength. Strength was slightly anisotropic (Figure 3) and we averaged tensile and shear strengths to 4.77 GPa and 4.00 GPa (Table 2), respectively. Interfacial failure strengths from nacre as a proxy-material (Table 2) were significantly lower than these single crystal strengths which indicates that interfacial rather than crystal failure dominates. Consequently, we used strengths reported for nacre under tension and shear (Table 2) as surrogates for the interfacial failure strengths at the aragonite needle level as initial values in our computations (Section 3.4).

### 3.3 Elasticity of the CWC skeleton not exposed to acidified waters

When increasing $\phi_{np}$, skeletal stiffness $\mathbb{S}_{CS}$ is significantly reduced (Figure 6a). Small changes in $\phi_{np}$ resulted in a linear decline of skeletal stiffness that became non-linear for porosities above 20% result which is illustrated in Figure 6 for Young's modulus and Poisson's ratio. Varying the aspect ratio while keeping $\phi_{np}$ fixed results in very subtle changes in $\mathbb{S}_{CS}$ for small porosities but more significant changes for higher porosities (Figure 6b). This compares well to results by Sanahuja et al. [38] and Fritsch et al. [57] and suggests that an aspect ratio of the aragonite crystals of 10 was a reasonable choice [32-35] for our purposes.

### 3.4 Micropillar compression tests of the skeleton not exposed to acidified waters

We obtained 144 usable micropillars on the 12 CWC samples (Section 2.7, Figure S4) with surface and base diameters of 28.96 µm (27.76-30.51 µm) and 86.5 µm (81.52-90.21 µm), a taper angle of 15.02° (12.16-18.19°), and a height of 110.77 µm (90.20-129.44 µm). We randomly selected six of the micropillars per sample for compression testing. Testing to failure resulted in a brittle response without damage accumulation (Figure 6d) with a median ultimate force of 517.86 mN (154.30-736.78 mN) and a median pillar stiffness of 487.9 mN/µm (225.02-1078.14 mN/µm). When separating these results into high and low aragonite concentration groups (Section 2.1), we obtain median ultimate forces of 516.40 mN (154.30-736.78 mN) and 517.86 mN (163.79-718.93 mN) that are not significantly different ($p \rightarrow 1$). However, we obtained median pillar stiffnesses of 633.41 mN/µm (225.02-1078.14 mN/µm) and 401.98 mN/µm (243.24-594.16 mN/µm) which are significantly lower for the low aragonite concentration group ($p < 10^{-12}$).

When interpreting the micropillar compression tests with our finite element model (Section S7) using the model predicted stiffness of a CWC skeleton with $\phi_{np} = 3.9\%$ (Figure 6) without further changes due to ocean acidification, we obtained a structural stiffness of the micropillar of 956.16 mN/µm. This is higher than the experimental median but within the range identified in the micropillar tests for healthy aragonite concentrations of 1.67-2.62. It is, however, higher than the results for samples from below the ASH (Figure 6d).

When using the polycrystalline yield criterion (Figure 4) with tensile and compressive strengths $\sigma_{poly}^{ut}$ and $\sigma_{poly}^{uc}$ based on $\sigma_{int}^{ut}$ and $\sigma_{int}^{sh}$ derived from nacre (Table 2) in our micropillar finite element model, we obtained an ultimate force of 289 mN. This is significantly lower than the experimentally achieved ultimate forces of 517.86 mN. Using the interfacial tensile and shear strengths for the aragonite single crystal would result in values over an order of magnitude larger. When increasing the interfacial yield strengths by a factor of 1.86 and keeping the ratio $\sigma_{int}^{ut}/\sigma_{int}^{us}$ similar to that of nacre, we obtained tensile and compressive polycrystal yield strengths (Table 2) that yield an ultimate force in our micropillar finite element model of 517.17 mN which is in good agreement with our experimental results. We therefore reject nacre as a surrogate material and consider the strengths derived by combining micropillar test results and the FE analyses as the active interfacial and polycrystalline strengths of the CWC skeleton (Table 2).

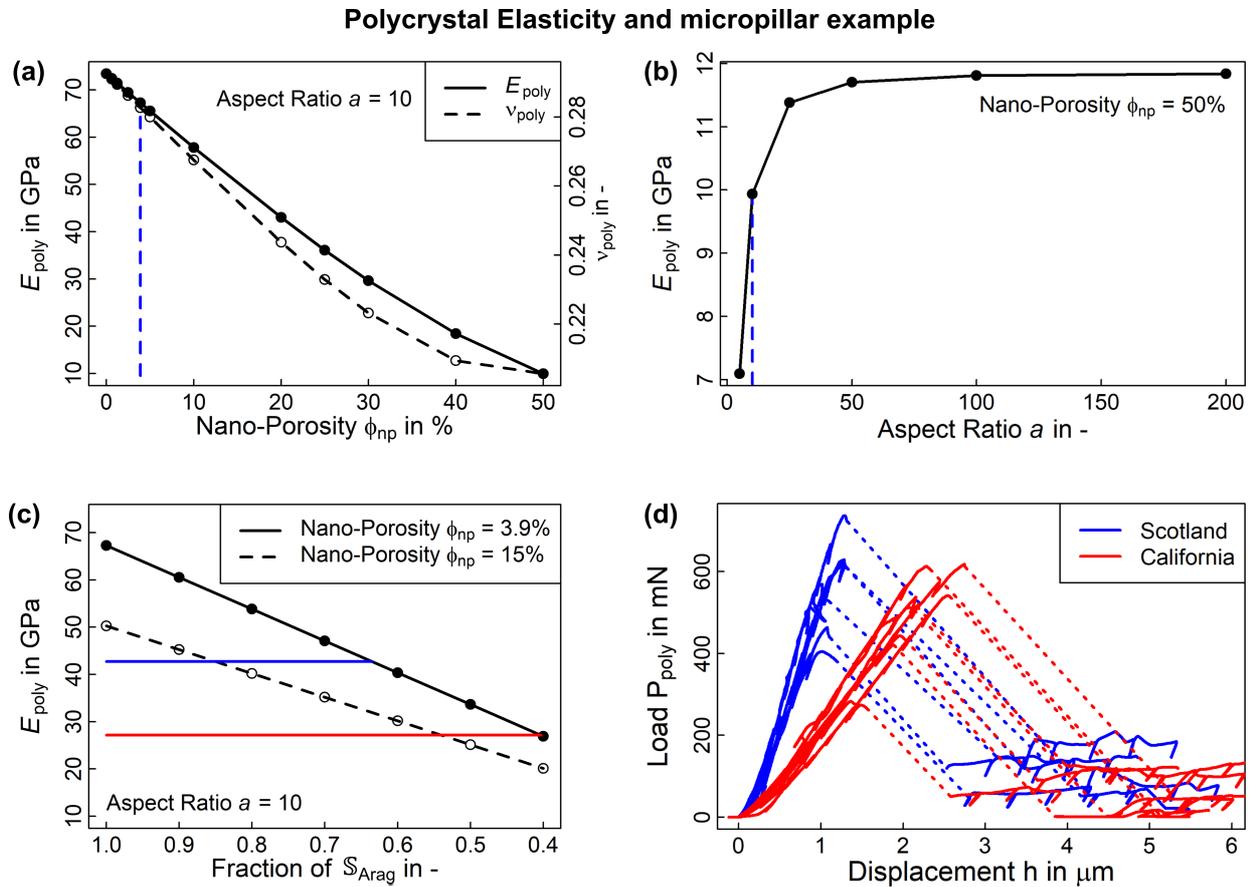

**Figure 6 Elasticity of the unaffected cold-water coral skeleton:** (a) Young's modulus $E_{poly}$ and Poisson's ratio $v_{poly}$ of the polycrystal depending on $\phi_{np}$ and shown for a crystal with an aspect of 10. The dashed blue line represents the chosen $\phi_{np} = 3.9\%$ for analysing strength. (b) illustrates the dependence of Young's modulus $E_{poly}$ on the crystal aspect ratio $a$ at $\phi_{np} = 50\%$. The dashed blue line illustrates the aspect ratio of 10 chosen for analysing strength. We illustrate dependence on the aspect ratio for $\phi_{np} = 50\%$ because for $\phi_{np} = 3.9\%$ Young's modulus varies $\pm\ 0.12$ GPa only. At $\phi_{np} = 50\%$, a higher aspect ratio than 10 could increase stiffness about 1.5 GPa. (c) Dependence of $E_{poly}$ on the stiffness of the aragonite crystals where we degraded $\mathbb{S}_{Arag}$ without changing its symmetry. Blue lines indicate medians of the Scottish (dashed) and California Sea Bight (dotted) samples. (d) compliance corrected results of six micropillar compression tests from two representative sample. Tests feature linear elasticity and a brittle behaviour after passing the ultimate point. The dashed line covers the zone of brittle failure that is expanded by compliance correction. See also Figure S4 for a SEM image of the micropillars.

## 3.5 Incorporating the impact of ocean acidification

Figure 7 illustrates the impact of ocean acidification induced porosity on the stiffness of the exposed CWC skeleton. At a maximum porosity of 33.3% as identified in our samples (Section 3.1), an affected volume of interest would suffer almost 50% loss of stiffness due to acidification. In contrast to *in situ* samples that show no porosity [9, 41, 42], this is a significant decrease in the ability of the skeleton to resist external loading. Likewise, an increase in porosity dramatically reduces strength of the CWC skeleton affected ocean acidification (Figure 7).

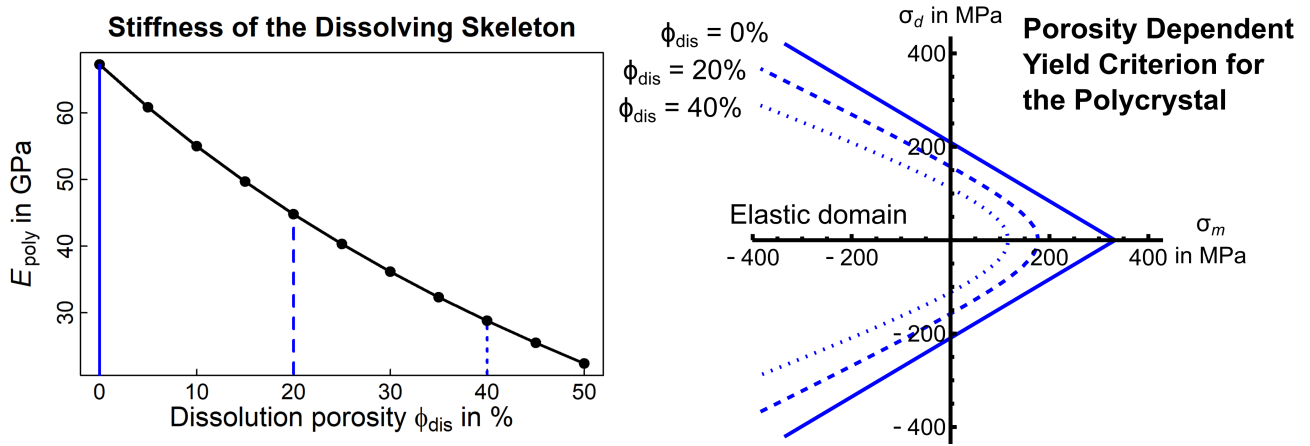

**Figure 7 Impact of ocean acidification on cold-water coral skeletal stiffness and strength:** Left, the effect of increasing porosity due to ocean acidification on the stiffness of the polycrystalline CWC skeleton. Small increases in porosity significantly lower the ability of the skeleton to resist external loading. Right, increasing porosity due to ocean acidification significantly lowers the limit bearable load the *dead* skeleton can withstand so that much lower loads are needed to break the CWC skeleton.

Using the developed material model in our image based finite element model of a representative CWC colony indicates that the degrading mechanism identified by Hennige et al. [9] can indeed be captured. Compliance of the structure is significantly increased either through an increase in porosity or through dissolution from the periphery of the skeletal wall (Figure 8). However, ocean acidification affects the *dead* coral skeleton through a combination of increased porosity and dissolution which leads to a decreased skeletal wall thickness. It is this combination that leads to the most detrimental increase in fragility (Figure 8).

## 4 Discussion

We here report the consequences of ocean acidification on the mechanical properties of CWCs using a multiscale material model complemented by experimental data (*in silico* and on physical samples). The developed model allows us to analyse failure of CWC colonies affected by ocean acidification over all relevant length scales. Strikingly, the synthesised skeletal building material is 10 times stronger than concrete [73] and twice as strong as ultrahigh performance fibre reinforced concrete [74] or nacre (Table 2). Our results indicate that the strength of the skeletal building material is retained even when skeletons are synthesised under future oceanic conditions (Figure 6, Section 3.4), as we demonstrate in samples from the California Sea Bight, a region that is considered to be representative of end-of-century oceanic conditions[9]. It is important to note the difference in length-scale here. The entire coral branch, apart from the live part[9], can be acidified and show porosity. However, the skeletal material of these coral branches is not porous but which was synthesised under different oceanic conditions and shows markable difference in crystal size. The novel multiscale mechanical properties identified here for the first time contradict the current state-of-knowledge that mineralised skeletons will generally weaken due to climate change. In fact, CWCs retain the strength of their building material despite a loss of its stiffness, and the threat comes rather from a loss of material as acidification induced increase in internal porosity and dissolution. Our multiscale mechanical model delivers a model-based explanation for this stunning feat.

### 4.1 Model verification and validation

To verify our model, we compare predicted stiffness with that derived from our micropillar tests via the associated micropillar finite element model and a set of independent nanoindentation results (Sections 3.4 and S8, Figure S6c). The ranges in polycrystalline stiffness measured by nanoindentation on *L. pertusa* samples from UK waters (Figure S6c), California Sea Bight [9], as well as non-zooxanthellate and zooxanthellate corals from the Mediterranean Sea [31, 41, 42] are in reasonably good agreement with our predicted polycrystalline stiffness with corals from the Mediterranean (red and green areas in comparison to the blue dashed line in Figure S6c). Interestingly, Pasquini et al. [31] reported homogeneous and isotropic nanoindentation results when testing in different directions which they relate to the microstructure. This supports the isotropy occurring for $\mathbb{S}_{CS}$ through choosing a self-consistent scheme, and implicitly verifies our modelling approach. When using the median transverse isotropic aragonite stiffness tensor, $\mathbb{S}_{Arag}$, our predicted stiffness agrees very well with the results by Pasquini et al. [31] (Figure 6 and S6c), although $\mathbb{S}_{CS}$ gives a higher stiffness than that derived by our nanoindentation results for *L. pertusa* from UK waters (Section S8, Figure S6c) and the California Sea Bight (grey area in Figure S6c). If the minimum aragonite stiffness tensor (Table 1) is used in $\mathbb{S}_{CS}$, polycrystalline stiffness derived here moves closer to the median nanoindentation stiffness which illustrates the influence of this variable on the overall outcome and needs further consideration.

In comparison to other studies [31, 41, 42], our nanoindentation results show (a) a high variation and (b) a lower median (Figure S6c). The large variation may be attributed to testing different crystal arrangements (sclerodermites [23]). This particularly affects the nanoindentation results, as probed volume is considerably smaller than that of our micropillar compression tests. When Hennige et al. [9] analysed the samples from California Sea Bight, no difference in compositional properties of the skeleton was found. This suggests that the variation (point (a)) is indeed caused by differences between individual crystal assemblies. When comparing nanoindentation results for *L. pertusa* samples from UK waters with those reported for samples from the California Sea Bight [9] there is no significant difference ($p = 0.08$), although median stiffness in Scottish samples is 8 GPa higher. Note, that we used a one-sided Wilcoxon signed rank test since a Shapiro-Wilk test and quantile-quantile plots rejected normality (Sec 2.10). Accepting that the idea of significance is gradation rather than a binary condition, the higher 8 GPa median in Scottish samples reflects our findings in the micropillar compression tests. With regards to the low median (point (b)), our nanoindentation testing protocol may have damaged the material underneath the tip and thereby lowered the median stiffness with a potentially greater effect on samples from the California Sea Bight that lived at much lower aragonite concentrations. Such damage may have also contributed to insignificant results when comparing nanoindentation results between samples from the California Sea Bight and UK waters.

This was confirmed by the micropillar tests whose median is closer to the higher values from the literature [31, 41, 42] and which show a lower variation (Section 3.4, Figure S6c). When evaluating these tests with the micropillar finite element model using our predicted stiffness tensor for the skeleton not affected by ocean acidification and without further modification, we overestimate the median structural stiffness of the micropillars, but are well within the identified range. The striking difference in comparison to our nanoindentation tests is that micropillar stiffness for the samples close to the ASH is significantly lower ($p < 10^{-12}$) while the maximum bearable load is not different ($p \to 1$). It is important to note that our samples were covered with soft-tissue so that they were not exposed to acidified waters [9] (Figure 1). To reach the stiffness of the unaffected skeleton for the low aragonite saturation samples, $\mathbb{S}_{CS}$ would need to include a $\phi_{np}$ between the crystals of ~20% (Figure 6). Hennige et al. [9] did not detect such a difference in $\phi_{np}$ and concluded that the skeleton above and below the ASH is similar and made of aragonite. A higher $\phi_{np}$ would also lower the strength of the polycrystalline material and, consequently, maximum bearable force which was not detected in our micropillar tests. $\phi_{np}$ on its own, therefore, does not provide a satisfying explanation for the detected difference in stiffness.

The identified interfacial strength (using $\mathbb{S}_{CS}$ with a $\phi_{np} = 3.9\%$, Figure 6a) at the crystal length scale is 1.86 times higher than was suggested when using nacre as a surrogate material (Figure 6c). It is, however, much closer to the interfacial strength of nacre than to the strength of the aragonite single crystal (Table 2). The interface, thus, dominates failure as in other polycrystals [37, 38]. While the polycrystal seems to be assembled similarly for low and high aragonite concentrations as long as crystals are there (Figure S1), the fact that the resulting polycrystalline strength is the same suggests that the crystal interface is crucial. Our results suggest that if polycrystalline stiffness decreases, interfacial strength is unaffected and polycrystalline strength is maintained. The lower stiffness but similar strength in the samples at low aragonite concentration is an exciting conundrum.

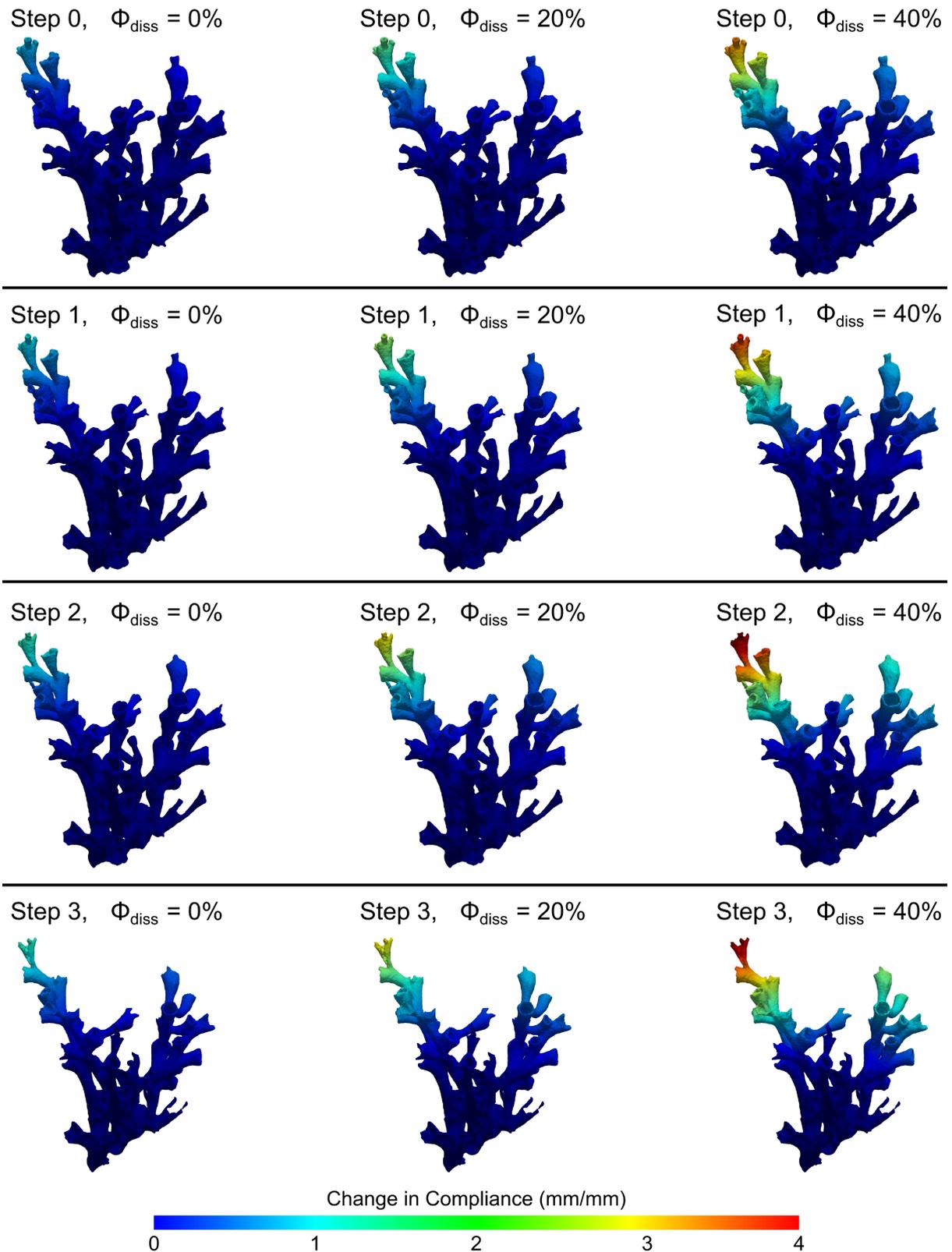

**Figure 8 Strength loss of an exemplary *L. pertusa* colony:** An increase in skeletal porosity (left to right) and an additional reduction in wall thickness (top to bottom) leads to decreasing load bearing capacity with the combination of both effects (bottom right) yielding a fourfold increase in flexibility resulting in a significantly increased fragility. For illustration purposes, we use deflection normalised on the top left case. The weaker the skeleton becomes the further peripheral elements can be deflected. Consequently, the highest internal loading is found towards the encastred base of the coral.

### 4.2 Retaining strength in a future ocean

Hennige et al. [9] detected significantly smaller crystals in samples from below the ASH without significant changes in $\phi_{np}$ or bulk density. This conundrum therefore requires other strategies to modify stiffness and strength of the multiscale material making up the skeleton. Interestingly, Kim et al. [75] found that including amino-acids in calcite (a polymorph of aragonite) can increase the hardness, which can be considered a proxy for strength, without affecting crystal stiffness. A similar effect was observed in supersaturated calcium-barium carbonate [76]. The mechanism behind this may be the creation of residual stresses by integration of intracrystalline organic or inorganic inclusions in a calcium carbonate that is crystallised from an amorphous precursor material [77]. In addition, smaller crystals have a higher surface to volume ratio so that contact area between them is increased. Such reinforcement may then stabilise the crystal interface so that interfacial cohesive forces increase, and polycrystalline strength remains the same. The reinforcement could be a compensation due to a lack of aragonite as a building material, which was revealed in smaller crystals overall [9]. The lowered stiffness may then either be caused by a breakdown in aspect ratio that can cause lower polycrystalline stiffnesses [38] or by lowering aragonite crystal stiffness (Figure 6c) while increasing load bearing capability to maintain interfacial and polycrystalline strength. In fact, when reducing aragonite stiffness by 30% and 50%, which is close to the median stiffness found in our micropillar tests (blue and red lines in Figure 6c), our models suggest that maintaining interfacial strength leads to insignificant changes in polycrystalline strengths (1.2% difference). This enables CWCs to cope with more compliant building materials while maintaining load bearing capacity potentially through utilisation of these hardening mechanisms [75-77]. This is corroborated by findings that skeletons formed under low aragonite concentrations show systematic crystallographic changes such as adapted crystal orientation and anisotropic distortion of the aragonite lattice [78] as well as suggestions that aragonite crystallography influences material properties such as strength [79]. Another mechanism to generate more compliant aragonite crystals could also result from crystallisation by attachment of amorphous particles, which Mass et al. [80] identified as the dominant crystallisation pathway, where flaws may be incorporated during accumulation of amorphous patches prior to proper crystallisation [81]. It seems however unlikely that the crystal stiffness is weakened by up to 60%. Re-interpreting the dashed black line in Figure 6c, we suggest that polycrystal stiffness is reduced by a combination of a slightly higher $\phi_{np}$, which may well have been below the detection limit of the techniques used by Hennige et al. [9], and a moderate reduction of aragonite stiffness.

Our results, therefore, suggest that CWCs have a way to deal with compromised availability of building material, and compensate resulting loss of stiffness through strengthening strategies that result in a significant increase of interfacial strength. This would then point to a conserved synthesis process of the polycrystal even in acidified waters. This also points to the robustness of our modelling as these effects can be explained by our stiffness and strength predictions through the properties at the crystal length scale. The proposed model can therefore capture the impact of crystallo-chemical changes on the multiscale mechanical behaviour of the reef structure [78]. We consider the proposed model usable to quantify elasticity and strength of the coral skeleton not exposed to corrosive waters and, more importantly, to investigate impact of exposure to acidified waters (Figure 1) and the mechanism suggested by Hennige et al. [9]. Finally, the smaller crystal size at low aragonite concentration may lead to a more crack tolerant material that is less damaged by nanoindentation than samples with larger crystals. This would provide an explanation why the difference in stiffness was not detected by nanoindentation since the potentially induced damage under tip was greater in samples from above the ASH and thereby lowering the stiffness more in these tests.

**4.3  Ocean acidification**

We used the model to test assertions by Hennige et al. [9] based on *in situ* and *in vitro* data that an increase in porosity and loss of material leads to a weakening of the *dead* coral foundation framework which consists primarily of exposed skeleton. For simplicity, we model acidification induced porosity in the whole skeleton and not only towards the outer periphery as identified by Hennige et al. [9]. This is achievable because these hollow, thin-walled structures are loaded primarily under bending, so that the critical stresses are located at the outer periphery (Figure 1d) with little effect of the internal periphery on the overall load bearing capacity. Increased fragility of such a weakened framework would lead to rapid crumbling of the overall reef structure and a reduction of the biodiversity supporting 3-dimensional complexity. Using a representative coral colony, Figure 8 illustrates that increasing porosity decreases loadbearing capacity of the skeleton and that this effect is worsened by an additional loss of material following dissolution [9, 16]. The loss of stiffness and strength already at small porosities (Figure 7) suggests that crumbling due to loss of load bearing capacity is a potentially rapid way in which these habitats will change. The relationship between porosity and aragonite concentration (Figure 1e, Hennige et al. [9]) provides an interesting opportunity, as porosity could be replaced by aragonite concentration as an independent variable. This would facilitate the use of a geochemical marker to quantify strength of the exposed skeleton *in situ* along with other variables such as temperature and oxygen concentration that would allow assessing health of the living part of CWCs reefs and, thus, provide means to monitor these sites locally.

It is of course conceivable that ocean acidification induced dissolution on ecosystem scales may be partially slowed by reaching equilibrium through dissolved aragonite even though the ASH is shoaling. In addition to acidification induced porosity, there is evidence [3] that future oceanic conditions with higher temperatures and lower oxygen may also lead to a significant increase in the efficacy of bioeroders. Breaking of coral under their own weight due to bioerosion are established but sporadic mechanisms to deteriorate CWC skeletons [82]. These bioeroders such as sponges, fungi, and borers create porosities that are either very similar [83] to those observed for low aragonite concentrations[9] or much larger, but with the same eventual mechanical impact. Tunnicliffe [84] was able to show the detrimental effect of increasing porosity on the mechanical competence of the coral skeleton by three-point bending testing. The mechanism was then used to explain breakage of colonies of *Acropora cervicornis* whose base was weakened by boring sponges [85]; very similar to the mechanism we described [9] (Figure 1) and which is captured by our model. This was corroborated by observations that borers decrease strength and increase fragility in massive corals with a dramatic loss of ability to withstand excessive loading caused by storms [86]. In addition, skeletal wall thinning and a limited uptake in porosity can be induced by Fe-Mn precipitated biofilms[87]. These films generate an edged appearance that is similar to the surface erosion Hennige et al. [9] report for their *in vitro* samples and promote attachment of other bioeroders that together result in similar weakening mechanisms – skeletal thinning and increased porosity. Therefore, irrespective of what generates porosity in a future ocean, the threat remains the same for CWC skeletons and the habitat they support.

Our modelling points to some crucial gaps of knowledge. Although we provide a robust estimate for the interfacial strength and an explanation of the variation in the results, our considerations with regards to the lower stiffness and similar maximum force needs experimental verification. The interfacial strength as suggested by nacre as a surrogate material proved too low and rules out a purely protein mediated assembly of the polycrystal. Our findings reject the assumption that dry nacre approximates the brittle, interfacial yield strength of aragonite [52] and it is an interesting question as to what generates crystal cohesion. Probably the most important gap is the absence of robust exposure times for CWCs under investigation. Hennige et al. [9] report the results shown in Figure 1e after 12 months experimental exposure in mesocosm experiments. However, *in situ* exposure time spans and robust markers to estimate exposure are currently missing. It is therefore not yet possible to use such a material model as a predictive tool to estimate risk of CWC loss. This would require further investigation on increasing porosities under different acidification scenarios to complement and solidify results of Hennige et al. [9] and *in situ* measurements of ocean chemistry. As most of the CWC reefs live currently above the ASH [1,7], such data would allow us to establish a baseline of aragonite concentrations with regular measurements allowing establishment of an exposure trajectory and thresholds that indicate time of exposure of CWCs to acidified waters.

### 4.4 Conclusion

We present a multiscale modelling framework that allows the investigation of the load bearing capacity of CWC structures. While our model underpins the dramatic and potentially rapid detrimental effects of ocean acidification to CWC skeletons it is an important step towards developing powerful monitoring tools. The impact of ocean acidification is dependent on the time CWCs are exposed to acidified waters. The model, therefore, allows us to investigate timescales of change as well as the impact of these changes on real reef structures if such an exposure time is known. It would therefore be possible to use the provided data to estimate time to reef crumbling, so that our results ultimately support future conservation and management efforts of these vulnerable marine ecosystems. It represents a crucial step towards understanding *which* ecosystems are at risk, *when* they will be at risk, and *how* much of an impact this will have upon associated biodiversity.

## Acknowledgements

This work was supported by a Leverhulme Trust Research Project Grant (RPG-2020-215) and the Engineering and Physical Sciences Research Council (EP/P005756/1) to UW, a 2021 student scholarship to ES by The Incorporation of Hammermen of Edinburgh, and by the Independent Research Fellowship from the Natural Environment Research Council (NERC) to SH (NE/K009028/1 and NE/K009028/2). North Atlantic corals used here were collected and experimentally treated through the UK Ocean Acidification programme (NE/H017305/1 to JMR). Californian samples were collected through support by NOAA National Centers for Coastal Ocean Science project "Vulnerability of Deep Sea Corals to Ocean Acidification", with additional funds from South Carolina Sea Grant Graduate Consortium R556, and the PADI Foundation 2013 Grant Award #7904. We would like to thank Scott Cavan, Jason Wakefield, and Guillaume Pinaqui for initial help with the micromechanical model and the finite element analyses.

## Author contributions

UW and SH conceptualised the study and solicited funding. AO contributed the molecular dynamics analyses. RB, SMcP, and JS performed laser ablation and associated SEM imaging. MPF, SMcP, and UW performed the micropillar compression tests and associated data analyses. SH contributed EBSD and synchrotron analyses. ES and MPF performed the image-based finite element analyses. CSE and UW conducted the porosity analyses. UW performed the nanoindentation analyses. JB and JT provided the computed tomography dataset of the exemplar sample. The manuscript was written by UW with contributions from MPF, AO, SMcP, ES, RB, MR, and SH. All co-authors assisted in reviewing and editing of the final manuscript.

## Declaration of Interest

The authors declare no conflict of interests.

# Supplementary material to:

# Multiscale Mechanical Consequences of Ocean Acidification for Cold-Water Corals

Uwe Wolfram[1*]   Marta Peña Fernández[1]   Samuel McPhee[1]   Ewan Smith[1]   Rainer J. Beck[1]   Jonathan D. Shephard[1]   Ali Ozel[1]   Craig Scott Erskine[1]   Janina Büscher[3]   Jürgen Titschack[4,5]   J Murray Roberts[2]   Sebastian Hennige[2]

[1]*School of Engineering and Physical Sciences, Institute of Mechanical, Process and Energy Engineering, Heriot-Watt University, Edinburgh, United Kingdom*
[2]*Changing Oceans Research Group, School of GeoSciences, University of Edinburgh, Edinburgh, United Kingdom*
[3]*GEOMAR Helmholtz Centre for Ocean Research Kiel, Biological Oceanography Research Group, Kiel, Germany*
[4]*Marum Center for Marine Sciences, University of Bremen, Bremen, Germany*
[5]*Senckenberg am Meer, Marine Research Department, Wilhelmshaven, Germany*
[*]*Corresponding author email: u.wolfram@hw.ac.uk*

## S1   Electron backscatter diffraction and scanning electron microscopy data

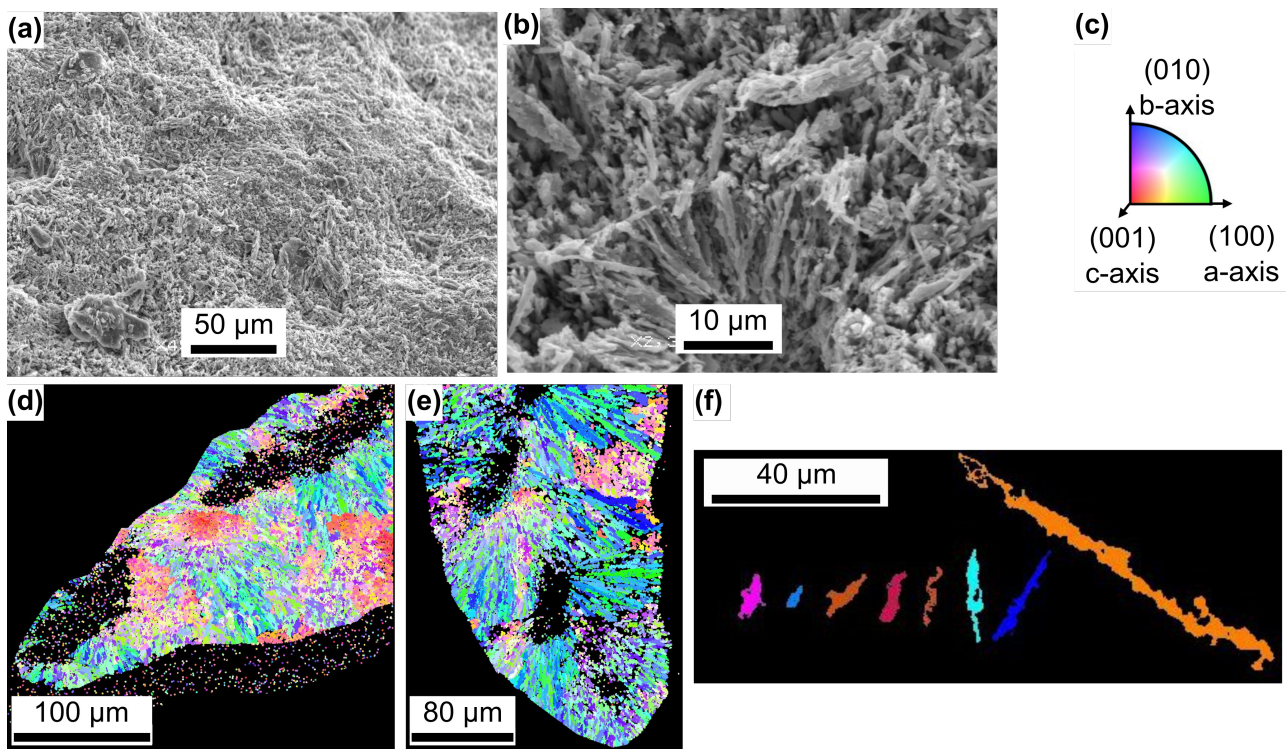

**Figure S1 Polycrystalline setup of *L. pertusa*:** SEM and EBSD data previously published by Hennige et al. [1] support that aragonite crystals coalesce in a polycrystalline matrix and that a random assembly is a suitable approximation. (**a, b**) are surface SEM images of cold-water coral skeletons. (**c**) shows the crystal coordinate systems used in the inverted pole figure image examples in (**d**) and (**e**) which were obtained using EBSD microscopy. EBSD data (**f**) shows eight examples across the 100 needles analysed which illustrate the range of aspect ratio of 2.01-13.85.

## S2 Porosity and affected layer analyses

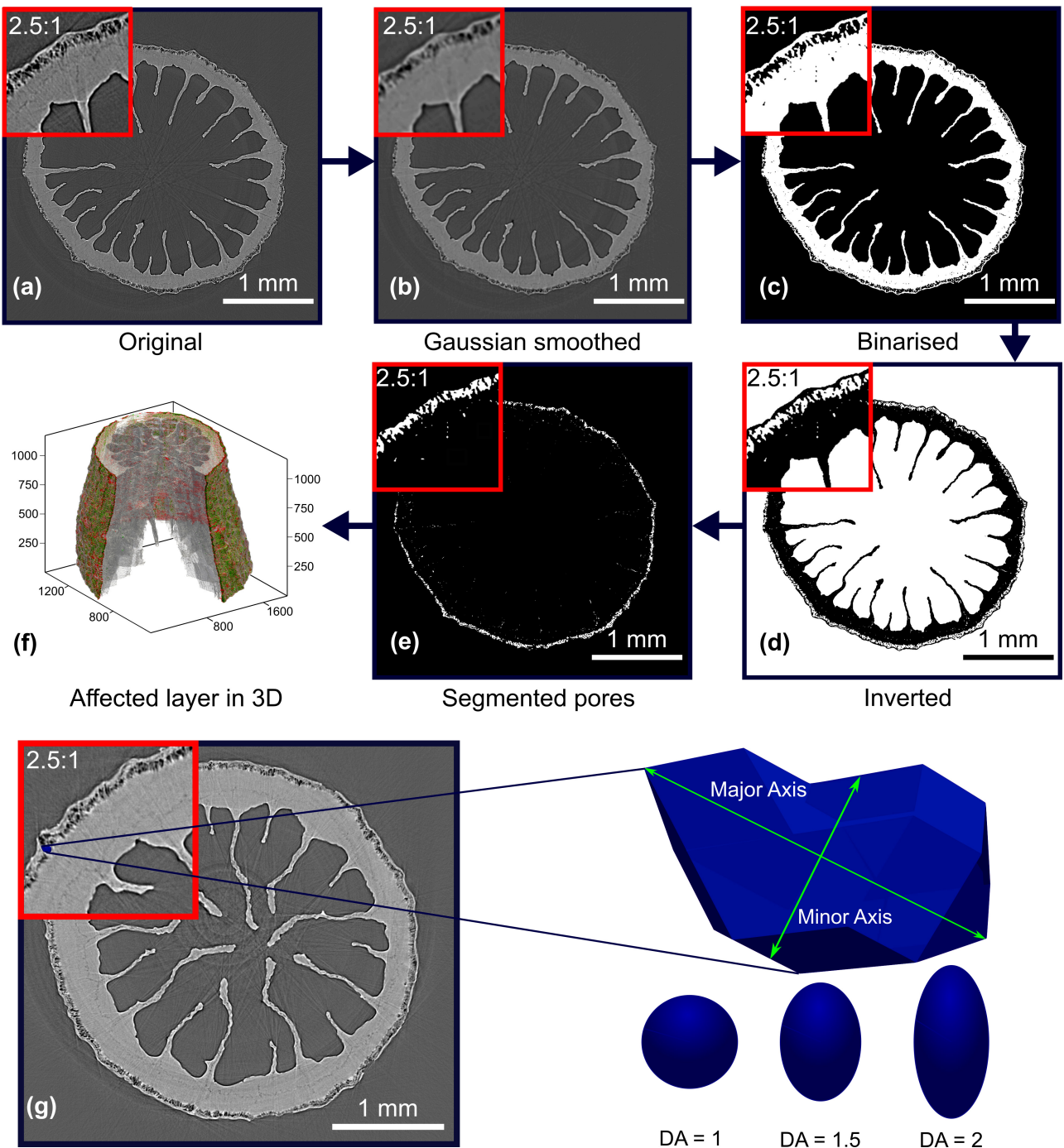

**Figure S2: (a-f)**: Sequential flowchart of porosity analysis. The original image was smoothed and thresholded to produce a binarised image. A mask was then generated using morphological operations (opening and closing). Pixel-wise multiplication of the inverse binarised and mask images enabled extraction of porosity. **(g)** Shape of individual pores were characterised by their Degree of Anisotropy (DA) using a connected component analysis. A larger DA value corresponds to a more oblong shape, while a sphere has a DA = 1.

## S3 Eshelby Tensors for Isotropic Inclusions

For an orthogonal material system, the 4th order Eshelby tensor Eshelby [2], $\mathbb{R}$, takes the form,

$$\mathbb{R} = \begin{pmatrix} R_{1111} & R_{1122} & R_{1133} & 0 & 0 & 0 \\ R_{2211} & R_{2222} & R_{2233} & 0 & 0 & 0 \\ R_{3311} & R_{3322} & R_{3333} & 0 & 0 & 0 \\ 0 & 0 & 0 & 2R_{2323} & 0 & 0 \\ 0 & 0 & 0 & 0 & 2R_{1313} & 0 \\ 0 & 0 & 0 & 0 & 0 & 2R_{1212} \end{pmatrix} \quad (S1)$$

The individual components of the Eshelby tensor for inclusion shapes used in this paper are provided below (solutions obtained from Mura [3] and David and Zimmerman [4]) where,

$v$ = Poisson's ratio of isotropic matrix
$a$ = Inclusion aspect ratio

$\mathbb{R}_{asi}$ - Aligned spheroidal inclusions

$$\mathbb{R}_{1111} = \mathbb{R}_{2222} = -\frac{3a^2}{8(1-v)(1-a^2)} + \frac{1}{4(1-v)}\left[1 - 2v + \frac{9}{4(1-a^2)}\right]g \quad (S2)$$

$$\mathbb{R}_{3333} = \frac{1}{1-v}\left[2 - v + \frac{1}{1-a^2}\right] + \frac{1}{2(1-v)}\left[-2(2-v) + \frac{3}{1-a^2}\right]g \quad (S3)$$

$$\mathbb{R}_{1122} = \mathbb{R}_{2211} = \frac{1}{8(1-v)}\left[1 - \frac{1}{1-a^2}\right] + \frac{1}{16(1-v)}\left[-4(1-2v) + \frac{3}{1-a^2}\right]g \quad (S4)$$

$$\mathbb{R}_{2233} = \mathbb{R}_{1133} = \frac{a^2}{2(1-v)(1-a^2)} + \frac{1}{4(1-v)}\left[1 - 2v + \frac{3a^2}{1-a^2}\right]g \quad (S5)$$

$$\mathbb{R}_{3311} = \mathbb{R}_{3322} = \frac{1}{2(1-v)}\left[-(1-2v) + \frac{1}{1-a^2}\right] + \frac{1}{4(1-v)}\left[2(1-2v) - \frac{3}{1-a^2}\right]g \quad (S6)$$

$$\mathbb{R}_{1212} = \mathbb{R}_{2121} = \frac{a^2}{8(1-v)(1-a^2)} + \frac{1}{16(1-v)}\left[4(1-2v) + \frac{3}{1-a^2}\right]g \quad (S7)$$

$$\mathbb{R}_{1313} = \mathbb{R}_{2323} = \frac{1}{4(1-v)}\left[1 - 2v + \frac{1+a^2}{1-a^2}\right] - \frac{1}{8(1-v)}\left[1 - 2v + 3\frac{1+a^2}{1-a^2}\right]g \quad (S8)$$

where $g$ is a function of the inclusion aspect ratio, $a$, and takes two forms depending on whether the inclusions are prolate spheroids ($a > 1$) or oblate spheroids ($a < 1$):

$$g = \frac{a}{(a^2-1)^{\frac{3}{2}}}\left(a(a^2-1)^{\frac{1}{2}} - \cosh^{-1} a\right) \quad \text{when } a > 1 \quad (S9)$$

$$g = \frac{a}{(1-a^2)^{\frac{3}{2}}}\left(\cos^{-1} a - a(1-a^2)^{\frac{1}{2}}\right) \quad \text{when } a < 1 \quad (S10)$$

$\mathbb{R}_{rsi}$ - Random spherical inclusions $(a_1 = a_2 = a_3 = a)$

$$\mathbb{R}_{1111} = \mathbb{R}_{2222} = \mathbb{R}_{3333} = \frac{7 - 5v}{15(1-v)} \quad (S11)$$

$$\mathbb{R}_{1212} = \mathbb{R}_{2323} = \mathbb{R}_{3131} = \frac{4 - 5v}{15(1-v)} \quad (S12)$$

$$\mathbb{R}_{1122} = \mathbb{R}_{2211} = \mathbb{R}_{3311} = \mathbb{R}_{1133} = \mathbb{R}_{2211} = \mathbb{R}_{3322} = \frac{5v - 1}{15(1-v)} \quad (S13)$$

## S4 Aragonite needle orientation

To find the most adversely loaded aragonite needle we employ needle coordinate systems introduced for randomly oriented hydroxyapatite needles[5, 6]:

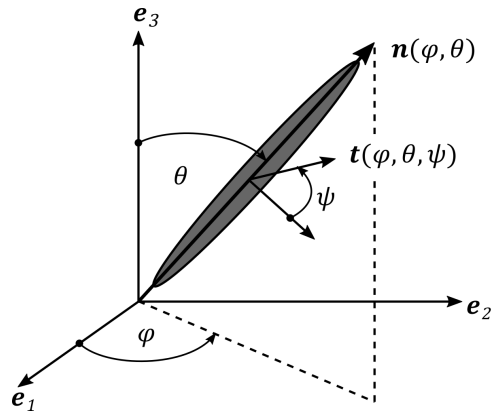

**Figure S3:** Crystal orientation characterised by Euler angles $\varphi, \theta$ and tangential vector $\boldsymbol{t}$ characterised by Euler angles $\varphi, \theta$ and the in-plane rotation $\psi$.

## S5 Micropillar compression

Two arrays of six micropillars were positioned on the skeletal wall of a CWC sample. Figure S4 illustrates micropillar location and shape on a representative sample.

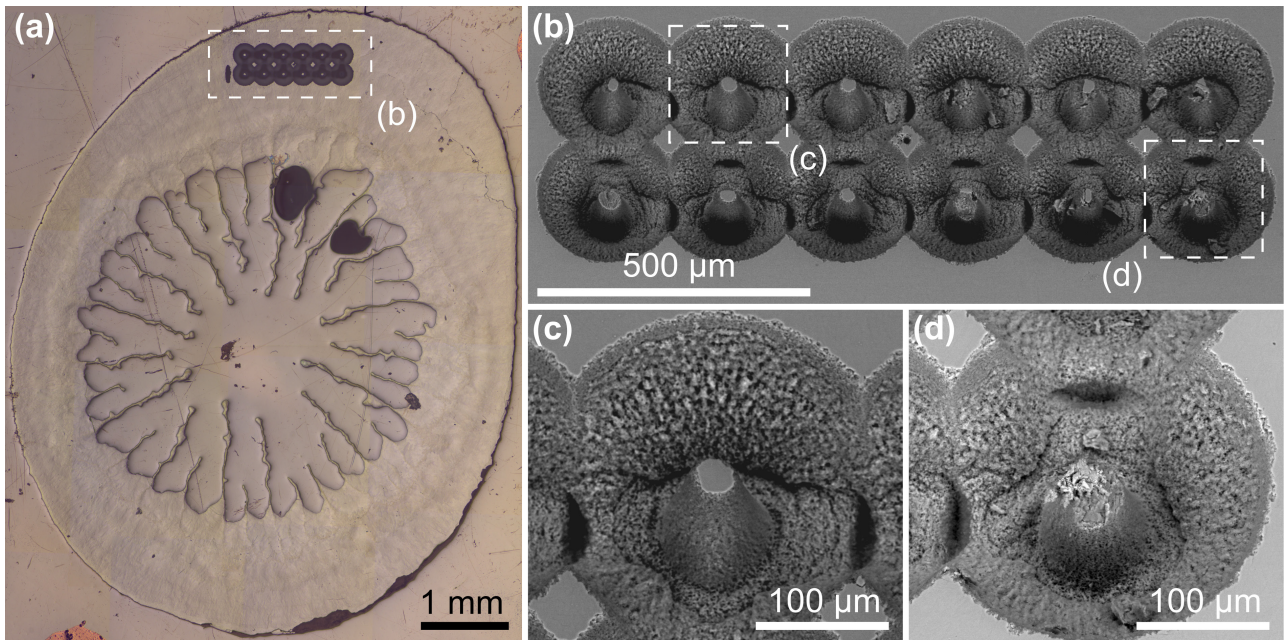

**Figure S4 Coral micropillars:** **(a)** reflected light microscopy image of a corallite cross section. **(b)** SEM image of a laser ablated array of 2x6 micropillars with the left six pillars before and the right six pillars after testing. **(c)** SEM of an untested micropillar and **(d)** SEM of a tested micropillar.

We fabricated micropillars using an ultrashort pulsed laser machining workstation based on a Carbide laser (Light Conversion) with a laser wavelength of $\lambda = 1028$ nm, a pulse length of 6 ps, and a repetition rate of 2 kHz. We used a focussed spot size of 20 μm, 89% beam overlap, and 11% spot separation. The laser was operated at a pulse energy of 10.3 μJ and the beam was brought to focus into a 20 μm diameter laser spot positioned 70 μm below the top surface of the coral tissue. The focussed laser beam was scanned across the sample in an inward Archimedean spiral pattern with a speed of 4.4 mm/s by means of a galvanometer scan head. The outer diameter of the spiral was 250 μm and the internal diameter 44 μm. This scanning pattern was repeated three times to create 2 x 6 pillar grids on each sample. After laser ablation, samples were cleaned in an ultrasound bath for 5 s to displace any material ejecta from the coral surface and then glued to SEM stubs for mounting.

## S6 Obtaining cohesion $h$ and friction coefficient $T$

The equations for $\mathbb{F}$ and $\mathbf{F}$ in equation (9) of the main manuscript are a porous version of a generalised Drucker-Prager criterion provided by Schwiedrzik et al.[7] with tensors:

$$\mathbb{F} = \frac{3}{2}F_0^2 \mathbf{I} \overline{\otimes} \mathbf{I} - \frac{1}{2}F_0^2 \mathbf{I} \otimes \mathbf{I} \quad \text{and} \quad \mathbf{F} = f_0 \mathbf{I} \quad \text{with} \quad F_0 = \frac{1}{2}\frac{\sigma^{uc}+\sigma^{ut}}{\sigma^{uc}\sigma^{ut}} \quad \text{and} \quad f_0 = \frac{1}{2}\frac{\sigma^{uc}-\sigma^{ut}}{\sigma^{uc}\sigma^{ut}}. \quad (S14)$$

Comparing (S1) with (9) for $\phi_{OA} = 0$ allows us to deduce $T$ and $h$ as:

$$T = \sqrt{6}\frac{f_0}{F_0} = \sqrt{6}\frac{\sigma^{uc}-\sigma^{ut}}{\sigma^{uc}+\sigma^{ut}} \quad \text{and} \quad h = \frac{1}{3f_0} = \frac{2}{3}\frac{\sigma^{uc}\sigma^{ut}}{\sigma^{uc}-\sigma^{ut}}. \quad (S15)$$

## S7 Elasto-viscoplastic material model and its implementation

We implemented this micromechanical model as a UMAT in Abaqus (6.16, Dassault Systèmes) following a similar material model for bone tissue[8]. We propose a linear elastic-viscoplastic material model whose elasticity is governed by our three-step micromechanical scheme (Figure 2). The elastic domain is limited by the Drucker-Prager type yield surface for the skeleton exposed to corrosive waters (Section 2.8). Motivated by 0.1-5wt% organic matter incorporated in the skeletal matrix as well as the creep behaviour in our nanoindentation experiments[9], we propose a viscoplastic post-yield behaviour (Figure S5). This allows us to adopt the model developed by Schwiedrzik and Zysset[8] for bone tissue. We present the governing equations since there are significant differences compared to Schwiedrzik and Zysset[8]. The observed mechanical behaviour shows no damage and features multiscale elasticity and a micromechanical yield surface with very short post-yield region under compression.

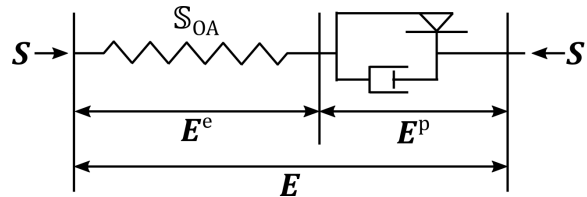

**Figure S5 Rheological model:** We propose a material model that consists of a linear-elastic spring and a short post-yield behaviour that is governed by a frictional slider and a dashpot to model viscoplastic behaviour. Viscosity here is motivated by 0.1-5wt% organic matter incorporated in the skeletal matrix.

We assume small deformations so that a Green-Naghdi split of total strain $\boldsymbol{E}$ into its elastic and plastic part is usable[10] and the accumulated plastic strain $\kappa$ can be defined as:

$$\boldsymbol{E} = \boldsymbol{E}^e + \boldsymbol{E}^p$$
$$\kappa = \int_0^t \left\|\dot{\boldsymbol{E}}^p\right\| d\tau. \tag{S16}$$

Using this, the free energy potential for this material takes on the form:

$$2\psi(\boldsymbol{E}, \boldsymbol{E}^p) = (\boldsymbol{E} - \boldsymbol{E}^p) : \mathbb{S}_{OA}(\boldsymbol{E} - \boldsymbol{E}^p) \tag{S17}$$

with $\mathbb{S}_{OA}$ the fourth order stiffness tensor for the acidified skeleton introduced in equation (8) of the main manuscript. Consequently, the stress is derived as

$$\boldsymbol{S} = \nabla_E \psi(\boldsymbol{E}, \boldsymbol{E}^p) = \mathbb{S}_{OA} : (\boldsymbol{E} - \boldsymbol{E}^p)$$
$$\boldsymbol{S}^p = -\nabla_{E^p} \psi(\boldsymbol{E}, \boldsymbol{E}^p) = \mathbb{S}_{OA} : (\boldsymbol{E} - \boldsymbol{E}^p) \tag{S18}$$

and the dissipation becomes

$$\Phi = \boldsymbol{S} : \dot{\boldsymbol{E}} - \dot{\psi} \geq 0$$
$$\Phi = \boldsymbol{S} : \dot{\boldsymbol{E}} - \boldsymbol{S} : \dot{\boldsymbol{E}} + \boldsymbol{S} : \dot{\boldsymbol{E}}^p = \boldsymbol{S} : \dot{\boldsymbol{E}}^p \geq 0. \tag{S19}$$

Motivated by 0.1-5wt% organic matter incorporated in the skeletal matrix as well as mild creep encountered in our nanoindentation experiments[9] we implement a viscoplasticity as proposed by Perzyna[8, 11, 12]:

$$\dot{\boldsymbol{E}}^p = \frac{1}{\eta} \langle \chi(Y_{OA}) \rangle \boldsymbol{M}^p$$
$$\boldsymbol{M}^p = \nabla_S Y_{OA}. \tag{S20}$$

$\langle ... \rangle$ represent the McAuley brackets, $Y_{OA}$ is the yield surface introduced in equation (9), and $\chi(Y_{OA})$ is a monotonously increasing and invertible overstress function necessary for Perzyna-type materials[8]. As proposed by Schwiedrzik and Zysset [8], we implemented a continuous Perzyna viscoplasticity through introducing a viscoplastic consistency parameter $\dot{\lambda}$ that we will substitute into the flow rule:

$$\dot{\lambda} = \frac{1}{\eta} \langle \chi(Y_{OA}) \rangle$$
$$\dot{\boldsymbol{E}}^p = \frac{1}{\eta} \dot{\lambda} \boldsymbol{M}^p \tag{S21}$$

We constrain the viscoplastic flow by exploiting invertibility of the overstress function so that we are able to formulate generalised Karush-Kuhn-Tucker conditions[8]:

$$Y_{OA} = \chi^{-1}(\dot{\lambda}\eta)$$
$$\bar{Y}_{OA} = Y_{OA} - \chi^{-1}(\dot{\lambda}\eta) = 0 \tag{S22}$$
$$\bar{Y}_{OA} \leq 0, \quad \dot{\lambda} \geq 0, \quad \dot{\lambda}\bar{Y}_{OA} = 0$$

This allows us then to implement the model as a UMAT in Abaqus (v6.16, Dassault Systémes). In the following, the increment number $n$ is omitted and converged variables $A_{n+1}$ will be called $A$ while variables at the beginning of the increment will be called $A_0$. The increment number is different from the local iteration number $i$ of the Newton–Raphson algorithm which we use to compute the unknowns $\boldsymbol{E}^p$ and $\kappa$. We start the computation with identifying a trial stress:

$$\boldsymbol{S}^T = \mathbb{S}_{OA}:(\boldsymbol{E} - \boldsymbol{E}_0^p) \tag{S23}$$

If the yield criterion is not violated by using this stress, i.e. $Y_{OA}(\boldsymbol{S}^T; \kappa_0) < 0$, the stress increment is elastic and the state variables can be updated as

$$\begin{aligned} \kappa &= \kappa_0 \\ \boldsymbol{E}^p &= \boldsymbol{E}_0^p \\ \boldsymbol{S} &= \boldsymbol{S}^T \end{aligned} \tag{S24}$$

and the tangent stiffness tensor is simply the elastic stiffness tensor $\mathbb{S}_{OA}$.

If $Y_{OA}(\boldsymbol{S}^T; \kappa_0) \geq 0$, an implicit stress return algorithm[13] is carried out respecting the Karush-Kuhn-Tucker conditions (S22.3) and the set of nonlinear equations to be solved is:

$$\begin{aligned} \boldsymbol{S} &= \mathbb{S}_{OA}:(\boldsymbol{E} - \boldsymbol{E}^p) \\ \bar{Y}_{OA}(\boldsymbol{S}, \kappa, \dot{\lambda}) &= Y_{OA}(\boldsymbol{S}, \kappa) - \chi^{-1}(\dot{\lambda}, \eta) = 0 \\ \dot{\boldsymbol{E}}^p &= \dot{\lambda} \nabla_{\boldsymbol{S}} Y_{OA}(\boldsymbol{S}, \kappa) \\ \dot{\kappa} &= \left\| \dot{\boldsymbol{E}}^p \right\| = \dot{\lambda} \left\| \nabla_{\boldsymbol{S}} Y_{OA}(\boldsymbol{S}, \kappa) \right\|. \end{aligned} \tag{S25}$$

The incremental Lagrangean multiplier $\Delta\lambda$ can be written as a function of the incremental accumulated plastic strain $\Delta\lambda = \frac{\Delta\kappa}{\|\nabla_{\boldsymbol{S}} Y_{OA}(\boldsymbol{S}, \kappa_0 + \Delta\kappa)\|}$ so that $\Delta\boldsymbol{E}^p = \Delta\kappa \frac{\nabla_{\boldsymbol{S}} Y_{OA}(\boldsymbol{S}, \kappa)}{\|\nabla_{\boldsymbol{S}} Y_{OA}(\boldsymbol{S}, \kappa_0 + \Delta\kappa)\|} = \Delta\kappa \boldsymbol{N}^p$. The polynomial flow rule for $\dot{\lambda}$ and the overstress function $\chi$ can be specified[8] as:

$$\begin{aligned} \dot{\lambda} &= \frac{1}{\eta}\left(Y_{OA}(\boldsymbol{S}, \kappa)^2 + m Y_{OA}(\boldsymbol{S}, \kappa)\right) \\ \chi^{-1}(\dot{\lambda}, \eta) &= -\frac{m}{2} + \sqrt{\frac{m^2}{4} + \eta\dot{\lambda}} \end{aligned} \tag{S26}$$

with $m = 1$ and $\eta = 1 \cdot 10^{-4}$ MPa·s.

Total strain at the end of the load increment is given and can be separated into $\boldsymbol{E} = \boldsymbol{E}_0 + \Delta\boldsymbol{E}$. With this the stress can be given in incremental form:

$$\boldsymbol{S} = \mathbb{S}_{OA}:(\boldsymbol{E}_0 + \Delta\boldsymbol{E} - \boldsymbol{E}_0^p - \Delta\boldsymbol{E}^p) = \boldsymbol{S}^T - \mathbb{S}_{OA}\Delta\boldsymbol{E}^p = \boldsymbol{S}^T - \mathbb{S}_{OA}\Delta\kappa\boldsymbol{N}^p \tag{S27}$$

wherein $\boldsymbol{S}^T = \mathbb{S}_{OA}:(\boldsymbol{E} - \boldsymbol{E}_0^p)$. We now multiply from the left with $\mathbb{E}_{OA} = \mathbb{S}_{OA}^{-1}$ and bring everything to one side to derive the residual error of the elastic strains

$$\boldsymbol{R}(\boldsymbol{S}, \Delta\kappa) = \mathbb{E}_{OA}(\boldsymbol{S} - \boldsymbol{S}^T) + \Delta\kappa \boldsymbol{N}^p. \tag{S28}$$

Similarly, we approximate the rate-dependent yield function by

$$\bar{Y}_{OA}(\boldsymbol{S}, \Delta\kappa) = Y_{OA}(\boldsymbol{S}, \kappa_0 + \Delta\kappa) - \chi^{-1}\left(\frac{\Delta\kappa}{||\nabla_{\boldsymbol{S}} Y_{OA}(\boldsymbol{S}, \kappa_0 + \Delta\kappa)||}\frac{1}{\Delta t}, \eta\right). \tag{S29}$$

(S28) and (S29) are linearised around the solution with respect to the variables $\boldsymbol{S}$ and $\Delta\kappa$ so that a linearised system of equations can be established:

$$\boldsymbol{R}^{i+1} = \boldsymbol{R}^i + \nabla_{\boldsymbol{S}}\boldsymbol{R}^i : \delta\boldsymbol{S} + \frac{\partial \boldsymbol{R}^i}{\Delta\kappa^i}\delta\Delta\kappa = \boldsymbol{0}$$
$$\bar{Y}_{OA}^{i+1} = \bar{Y}_{OA}^i + \nabla_{\boldsymbol{S}}\bar{Y}_{OA}^i \delta\boldsymbol{S} + \frac{\partial \bar{Y}_{OA}^i}{\Delta\kappa^i}\delta\Delta\kappa = 0 \tag{S30}$$

This is then solved iteratively using $\mathbb{S}_a = -(\nabla_{\boldsymbol{S}}\boldsymbol{R})^{-1}$ and

$$\delta\Delta\kappa = -\frac{\frac{\bar{Y}_{OA}^i}{||\nabla_{\boldsymbol{S}}\bar{Y}_{OA}^i||} + \boldsymbol{N}^p \mathbb{S}_a \boldsymbol{R}^i}{\boldsymbol{N}^p \mathbb{S}_a \frac{\partial \boldsymbol{R}^i}{\partial \Delta\kappa^i} + \frac{\frac{\partial \bar{Y}_{OA}^i}{\partial \Delta\kappa^i}}{||\nabla_{\boldsymbol{S}}\bar{Y}_{OA}^i||}} \quad \text{and} \quad \delta\boldsymbol{S} = \mathbb{S}_a\left(\boldsymbol{R}^i + \frac{\partial \boldsymbol{R}^i}{\Delta\kappa^i}\delta\Delta\kappa\right) \tag{S31}$$

until $||\boldsymbol{R}|| < tol$ and $Y < tol$ wherein $tol$ is a predefined tolerance. The solution in each iteration is used to update the current state of the variables:

$$\boldsymbol{S}^{i+1} = \boldsymbol{S}^i + \delta\boldsymbol{S}$$
$$\Delta\kappa^{i+1} = \Delta\kappa^i + \delta\Delta\kappa \tag{S32}$$

Once converged, the state variables can be updated so that $\boldsymbol{S} = \boldsymbol{S}^{i+1}$, $\kappa = \kappa_0 + \Delta\kappa^{i+1}$, and $\boldsymbol{E}^p = \boldsymbol{E} - \mathbb{E}_{OA}\boldsymbol{S}$.

According to Simo and Hughes [13] the algorithmic tangent tensor $\mathbb{S}_{CA}$ can be found through a linearisation of the stress-strain relationship around the current solution and by enforcing consistency. Schwiedrzik and Zysset [8] conclude that this is already done during application of the Newton-Raphson scheme for stress integration. Therefore, the tensor relating infinitesimal changes in strain $\delta\boldsymbol{R}$ to infinitesimal changes in stress $\delta\boldsymbol{S}$ (i.e. the tangent operator) can be found by substituting (S31.1) at $Y_{OA} = 0$ into (S31.2):

$$\delta\boldsymbol{S} = \mathbb{S}_a \boldsymbol{R}^i - \mathbb{S}_a \frac{\partial \boldsymbol{R}^i}{\Delta\kappa^i}\left(\frac{\boldsymbol{N}^p \mathbb{S}_a \boldsymbol{R}^i}{\boldsymbol{N}^p \mathbb{S}_a \frac{\partial \boldsymbol{R}^i}{\partial \Delta\kappa^i} + \frac{\frac{\partial \bar{Y}_{OA}^i}{\partial \Delta\kappa^i}}{||\nabla_{\boldsymbol{S}}\bar{Y}_{OA}^i||}}\right) = \left(\mathbb{S}_a - \frac{\mathbb{S}_a\left(\frac{\partial \boldsymbol{R}^i}{\partial \Delta\kappa^i} \otimes \boldsymbol{N}^p\right)\mathbb{S}_a}{\boldsymbol{N}^p \mathbb{S}_a \frac{\partial \boldsymbol{R}^i}{\partial \Delta\kappa^i} + \frac{\frac{\partial \bar{Y}_{OA}^i}{\partial \Delta\kappa^i}}{||\nabla_{\boldsymbol{S}}\bar{Y}_{OA}^i||}}\right)\boldsymbol{R}^i \tag{S33}$$

$$\delta\boldsymbol{S} = \mathbb{S}_{CA}\boldsymbol{R}^i$$

The derivatives used in (S30) to (S33) amount to:

$$\nabla_S \boldsymbol{R}^i = \mathbb{E}_{OA} + \Delta\kappa^i \nabla_S \boldsymbol{N}^p$$

$$\frac{\partial \boldsymbol{R}^i}{\partial \Delta\kappa^i} = \boldsymbol{N}^p + \Delta\kappa^i \frac{\partial \boldsymbol{N}^p}{\partial \Delta\kappa^i} \tag{S34}$$

$$\nabla_S \boldsymbol{N}^p = \frac{\nabla_S(\nabla_S Y_{OA}^i) ||\nabla_S Y_{OA}^i|| - \nabla_S Y_{OA}^i \otimes \left(\nabla_S ||\nabla_S Y_{OA}^i||\right)}{||\nabla_S Y_{OA}^i||^2}. \tag{S35}$$

For the rate independent yield surface, we introduce a hardening function $r(\kappa)$ that allows to take different forms of post-yield behaviour such as exponential hardening or softening into account so that the yield surface and associated derivatives become:

$$Y_{OA}(\boldsymbol{S}, \Delta\kappa^i) \coloneqq \sqrt{\boldsymbol{S}:\mathbb{F}\boldsymbol{S}} + \boldsymbol{F}:\boldsymbol{S} - r(\kappa_0 + \Delta\kappa^i) = 0 \tag{S36}$$

$$\frac{\partial Y_{OA}^i}{\partial \Delta\kappa^i} = -r'(\Delta\kappa^i) \quad \text{and} \quad \nabla_S Y_{OA}^i = \frac{\mathbb{F}:\boldsymbol{S}}{\sqrt{\boldsymbol{S}:\mathbb{F}\boldsymbol{S}}} + \boldsymbol{F} \tag{S37}$$

$$\frac{\partial ||\nabla_S Y_{OA}^i||}{\partial \Delta\kappa^i} = 0 \quad \text{and} \quad \frac{\partial \nabla_S Y_{OA}^i}{\partial \Delta\kappa^i} = 0 \tag{S38}$$

$$\nabla_S(\nabla_S Y_{OA}^i) = \frac{\mathbb{F}(\boldsymbol{S}:\mathbb{F}\boldsymbol{S}) - (\boldsymbol{S}:\mathbb{F})^T \otimes (\mathbb{F}:\boldsymbol{S})}{(\boldsymbol{S}:\mathbb{F}\boldsymbol{S})^{\frac{3}{2}}}$$

$$\nabla_S ||\nabla_S Y_{OA}^i|| = \frac{\nabla_S(\nabla_S Y_{OA}^i):\nabla_S Y_{OA}^i + \nabla_S Y_{OA}^i:(\nabla_S Y_{OA}^i)}{||\nabla_S Y_{OA}^i||}. \tag{S39}$$

The derivatives for the rate dependent yield surface are then a combination of the rate independent case and the viscous correction:

$$\bar{Y}_{OA}(\boldsymbol{S}, \Delta\kappa) = Y_{OA}(\boldsymbol{S}, \kappa_0 + \Delta\kappa^i) + \frac{m}{2} - \sqrt{\frac{m^2}{4} + \frac{\eta}{\Delta t} \frac{\Delta\kappa^i}{||\nabla_S Y_{OA}(\boldsymbol{S}, \kappa_0 + \Delta\kappa^i)||}} \tag{S40}$$

$$\bar{Y}_{OA}^i = Y_{OA}^i + \frac{m}{2} - \left(\frac{m^2}{4} + \frac{\eta}{\Delta t}\frac{\Delta\kappa^i}{||\nabla_S Y_{OA}^i||}\right)^{\frac{1}{2}}$$

$$\nabla_S \bar{Y}_{OA}^i = \nabla_S Y_{OA}^i + \frac{1}{2}\left(\frac{m^2}{4} + \frac{\eta}{\Delta t}\frac{\Delta\kappa^i}{||\nabla_S Y_{OA}^i||}\right)^{-\frac{1}{2}} \frac{\eta}{\Delta t} \frac{\nabla_S ||\nabla_S Y_{OA}^i||}{||\nabla_S Y_{OA}^i||^2}\Delta\kappa^i \tag{S41}$$

$$\frac{\partial \bar{Y}_{OA}^i}{\partial \Delta\kappa^i} = \frac{\partial Y_{OA}^i}{\partial \Delta\kappa^i} + \frac{1}{2}\left(\frac{m^2}{4} + \frac{\eta\Delta\kappa^i}{\Delta t ||\nabla_S Y_{OA}^i||}\right)^{-\frac{1}{2}} \frac{\eta}{\Delta t}\frac{1}{||\nabla_S Y_{OA}^i||}$$

This material model was used to interpret the micropillar compression tests (Sections 3.7, 3.4) and to simulate CWC deformation due to a distributed pressure (Section 3.5, Figure S6).

**S7.1  Interpreting micropillar compression tests**

To interpret the micropillar compression tests, pillar dimensions were based upon the median dimensions of the manufactured micropillars (Sections 2.7 and 3.4) with a top radius of 14.48 µm, 43.17 µm radius for the base, and 110.77 µm pillar height. Substrate material was modelled with a radius and length of 150 µm to accommodate "sink-in" of the pillar as it is compressed. One quarter of the pillar and substrate was modelled with symmetry constraints in the x-z and y-z planes applied accordingly (Figure S6). The bottom surface of the substate was constrained against displacement in the loading testing direction (y-axis). The microindenter was not modelled but represented through a displacement boundary condition applied to the top surface. Displacement was set to 3 µm, enough to pass the yield point of the material. The entire model was meshed with quadratic tetrahedral elements (C3D10, Figure S6). Due to the complexity of the stress field under the pillar base, meshing was constrained by a single bias seed to maintain a fine mesh, while the substrate's mesh seed is biased to reduce in size as it approaches the edges of the model to save element count and thus computation cost. A mesh sensitivity analyses was conducted to evaluate at which point the recorded reaction force saturates and does not change with further mesh refinement. For this pillar this was found to be at 454,786 elements. Force-displacement data was retreived from the model by recording the displacement of the pillar's top surface and the cummulative reaction force upon every node on the substrate base.

**S7.2  Simulating CWC deformation due to a distributed pressure**

Surface and volumetric meshes (Section 2.9) were imported into Matlab (R2020a), where input files for Abaqus (v6.16, Dassault Systémes) were generated. The coral skeleton surface was subjected to a constant pressure load simulating an arbitrarily chosen sea current with a velocity of 3 m/s in a direction perpendicular to the longitudinal axis (z-axis) of the skeleton (Figure S6). This velocity was chosen to firmly overload the sample and represents three times the maximum water current reported by Haugan et al. [14] (0.3-1 m/s). It is important to note that this is an academic example to illustrate the effect of increasing velocity and loss of skeletal wall thickness. First, the normal of all external faces were computed and those facing the flow direction identified. A ray-triangle intersection algorithm[15] was then used to calculate the intersection of a ray in the direction of the flow and the triangulated surface mesh. This allowed us to detect those faces not shielded by other elements, thus, isolating the triangular faces on the coral skeleton surface where the rays first impinge on. The tetrahedral elements to which such faces belonged were identified and a distributed surface pressure $p$ was applied following $p = {1}/{2}\,\rho v^2 \cos\alpha$. $\rho$ is the density of the sea water ($\rho = 1026$ kg m$^{-3}$), $v$ the velocity of the current, and $\alpha$ the angle between the direction of the sea current flow and the normal of the faces hit by the flow (Figure S6). We omitted shear forces and the hydrostatic pressure for simplicity knowing that they would contribute to the actual mechanical situation. Considering flow in our example was sufficient to illustrate the impact of ocean acidification on load bearing capacity. Finally, linear tetrahedral elements were converted to quadratic ones and all degrees of freedom in the nodes of the most distal portion of the skeleton (10 mm height) were constrained to simulate a coral fixed at its root.

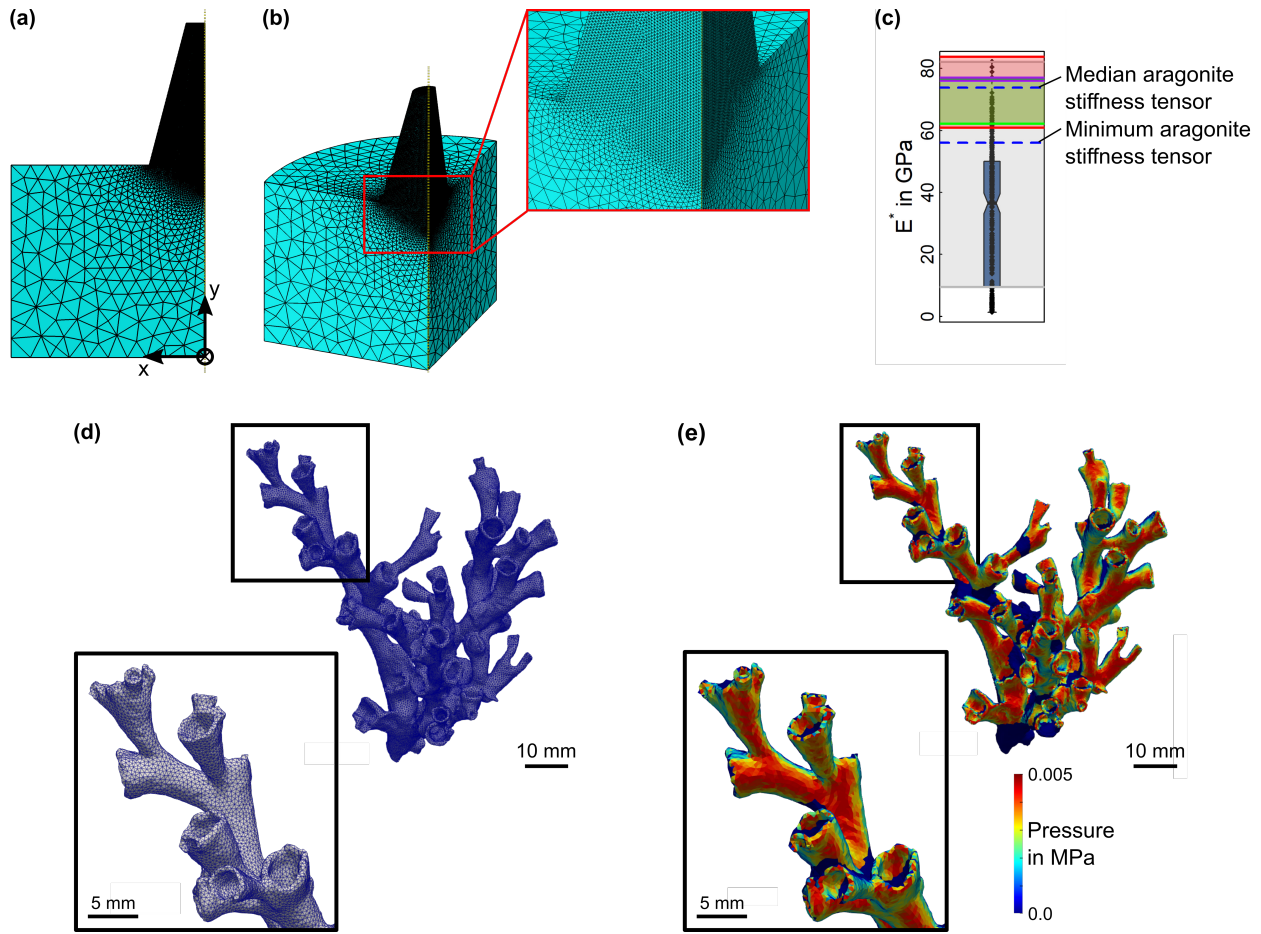

**Figure S6 Finite element (FE) models used in the study:** **(a)** and **(b)** finite element mesh of the fabricated micropillars to identify compressive strength of the polycrystalline material (Sections 2.7 and 3.4). **(c)** Stiffness obtained by nanoindentation tests (Section S8). The areas represent ranges for stiffness found in the literature (grey: California Sea Bight[9], red: shallow water corals around volcanic sites in the Mediterranean Sea[16], green: non-zooxanthellate and zooxanthellate corals from the Mediterranean Sea[17]; purple: scleractinian corals from Mediterranean and tropical waters[18]) The dashed blue lines **(c)** illustrate the results from our modelling (Section 3.3). Our model prediction using median aragonite stiffness fits very well to results by Pasquini et al. [18]. **(d)** Tetrahedral mesh of a representative coral sample that was used to investigate the impact of ocean acidification. **(e)** pressure boundary conditions loading the coral with a distributed flow pressure.

## S8  Measuring skeletal stiffness using nanoindentation

Hennige et al. [9] measured skeletal stiffnesses of *L. pertusa* samples from the California Sea Bight which were collected along an aragonite saturation range of 0.71-1.11. Samples were grouped in *live* and of *dead* coral skeletons (skeletons no longer covered in soft tissue) and no difference between the two groups were found. We extend this dataset by including results of 15 *L. pertusa* samples collected from UK waters (three samples from Mingulay Reef, Rockall Bank, Logachev Mound, Pisces 9, and Porcupine Seabight each)[1] which were covered with living tissue at collection. Samples were collected along an aragonite saturation range of 1.67-2.62 and we consider them to be representative of a non-acidified oceanic environment.

Sample preparation was the same as in Hennige et al. [9] and followed previously developed protocols for mineralised tissues[19, 20]. Testing protocol was also kept the same as in Hennige et al. [9]. Briefly, indentations were performed in dry conditions using a Berkovich tip mounted to a depth-sensing, force controlled nanoindenter (Hysitron). Force was applied in a monotonic ramp up to 50 mN over the course of 60 s. Subsequently, force was held constant for 30 s before being unloaded in 7.5 s. 35 indentations per sample (525 indentations in total) were performed and plain strain modulus, hardness, as well as the ratio between elastic and dissipated work were determined following Hennige et al. [9], with plain strain modulus being the interesting variable for validating our polycrystalline modelling. The resulting indentation stiffness is shown in Figure S6c.